\documentclass{emulateapj}

\lefthead{Bekki}
\righthead{The Galaxy formation}

\begin{document}

\title{Origin of chemical and dynamical properties of the Galactic thick disk}

\author{Kenji Bekki} 
\affil{
ICRAR,
M468,
The University of Western Australia
35 Stirling Highway, Crawley
Western Australia, 6009, Australia
}
\author{Takuji Tsujimoto}
\affil{
National Astronomical Observatory, Mitaka-shi, Tokyo 181-8588, Japan}

\begin{abstract}
We adopt a scenario  in which the Galactic thick disk was formed by minor merging
between the first generation of the Galactic thin disk (FGTD) and a dwarf galaxy
about $\sim 9$ Gyr ago
and thereby investigate chemical and dynamical properties of the Galactic thick disk.
In this scenario, the dynamical properties of the thick disk have long been
influenced both by the mass growth of the second generation
of the Galactic thin disk (i.e., the present thin disk)  and by
its non-axisymmetric structures. On the other hand,  the early 
star formation history and chemical
evolution of the thin disk was influenced by the remaining gas of the thick disk.
Based on N-body simulations and chemical evolution models,
we investigate the radial metallicity gradient, 
structural and kinematical properties,
and detailed chemical abundance patterns
of the thick disk.
Our numerical simulations show that  the ancient minor merger
event can significantly flatten the original radial metallicity gradient
of  the FGTD, in particular, in the outer part, and also can be responsible for
migration of inner metal-rich stars into the outer part ($R>10$kpc).
The simulations show that the central region
of the thick disk  can develop a bar 
due to dynamical effects of a separate  bar in the thin disk.
Whether  rotational
velocities ($V_{\phi}$) can correlate with
metallicities ([Fe/H]) for the simulated thick disks
depends on 
the initial metallicity
gradients of the FGTDs.
The simulated  orbital eccentricity distributions 
in the thick disk for models with higher mass-ratios ($\sim 0.2$) 
and lower orbital eccentricities ($\sim 0.5$) of 
minor mergers are in good agreement
with the corresponding observations.
The simulated $V_{\phi}$-$|z|$ relation of the thick disk 
in models with low orbital inclination angles of mergers are also in good agreement
with the latest observational results. The vertical metallicity gradient of the simulated thick 
disk is rather flat or very weakly negative at the solar neighborhood.
Our Galactic chemical evolution models 
show that if we choose two distinctive timescales for star formation
in the thin and thick disks,
then the models can explain both the observed 
metallicity distribution functions (MDFs) and
correlations between  [Mg/Fe] and [Fe/H] for the two disks 
in a self-consistent manner.
We discuss how the early star formation history and chemical evolution
of the Galactic thin disk can be influenced by the pre-existing thick disk.  
\end{abstract}

\section{Introduction}

The thick disk of the Galaxy is a  fundamental Galactic component 
containing  fossil records of  the early chemical and dynamical
evolution of the Galaxy 
(e.g., Freeman 1987; Majewski 1993; Freeman \& Bland-Hawthorn 2002).
Since Yoshii (1982) and Gilmore \& Reid (1983) first revealed the presence of
the thick disk
with scale height $\sim 1350$pc,  the characteristic structural, kinematical,
and chemical properties have been discussed extensively  in both 
observational and theoretical studies 
(e.g., Yamagata \& Yoshii  1992; Robin et al. 1996;
Gilmore et al. 2002; Reddy 2010).  
The  older  age  (e.g., Gilmore et  al. 1995; Bensby et al. 2003), 
higher [$\alpha$/Fe] (e.g., Fuhrmann 1998), larger scale-height 
(e.g., Ojha 2001, Larsen \& Humphresy 2003; Juri\'c et al. 2008),
and larger radial,  azimuthal, and vertical velocity dispersions  
(e.g., Soubiran et al. 2003; Dinescu et al. 2011) of the thick disk
in comparison with the thin disk
have long been considered to provide  
constraints on the formation models of the thick disk.
Theoretical models for the formation of the Galaxy  based on a cold dark matter   
cosmology have  started to explore the origin of
the observed complicated nature of the thick disk
(e.g., Abadi et al. 2003; Brook et al. 2004).

Previous numerical simulations showed that accretion of a small satellite
galaxy on  the pre-existing thin stellar 
disk of a disk galaxy can dynamically heat up
the disk to form the thick disk (e.g., Quinn et al. 1993; 
Walker et al. 1996; Huang \& Carlberg 1997; 
Velazquez \& White 1999 Villalobos \& Helmi 2008):
this formation process is referred to as the ``minor merger'' scenario.
Using cosmological simulations with self-consistent chemical evolution
models,
Brook et al. (2004) demonstrated that thick disks of galaxies can be formed
by accretion/merging  of gas-rich dwarfs 
at high redshifts
(this is referred to as the ``gas-rich merger'' scenario). 
Radial migration of stars due to internal stellar dynamics of disk galaxies
(Sellwood \& Binney 2002) has been demonstrated to be important for the formation
of the Galactic thick disk (e.g., Sch\"onrich \& Binney 2009; Loebman et al. 2010). 
Noguchi (1998) first showed that
massive clumps developed in the early dynamical evolution of galactic disks
can heat up the original disks to form thick disks.

The predicted properties of the Galactic thick disk
from these  formation scenarios have recently been compared with 
the present-day properties of the thick disk:
the orbital 
eccentricity distribution
(e.g., Sales et al. 2009; Dierickx et al. 2010; Dinescu et al. 2011),
radial and vertical metallicity gradients (e.g., Allende Prieto et al. 2006; 
Ivezi\'c et al. 2008),  [$\alpha$/Fe]-[Fe/H] relation different from
that of the thin disk (e.g., Reddy \& Lambert 2009; Reddy 2010), 
and rotational velocities of thick disk stars
dependent on vertical distances   (e.g., Allende Prieto et al. 2006)  and on
metallicities (Spagna et al. 2010).
These comparisons between theoretical and observational results have pointed out
strength and weakness  of the above mentioned scenarios in explaining
chemical and dynamical properties of the Galactic thick disk. For example,
the observed orbital eccentricity distribution of the thick disk stars
is consistent either with the gas-rich merger scenario
or the minor merger scenario (e.g., Dierickx et al. 2010; Di Matteo et al. 2011 Dinescu
et al. 2011; Wilson et al. 2011).
These comparisons however are not  yet extensive enough to
decide which of the proposed scenarios  is the most self-consistent
and convincing.

Although previous numerical studies based on the minor merger
scenario provided some key predictions on 
structural and kinematical properties of the Galactic thick disk
(Quinn et al. 1993;
Walker et al. 1996; Huang \& Carlberg 1997; Villalobos \& Helmi 2008),
they did not discuss chemical properties nor their correlation with the
dynamical properties observed in the Galactic thick disk. 
Given that the thick disk is composed largely of rather old stars 
with ages larger than $\sim 9$ Gyr (e.g., Allende Prieto et al. 2006;
Bensby et al. 2007), 
the formation epoch of the thick disk should be  well before the formation
of the thin disk,  and thus dynamical properties of the thick disk should
have been strongly influenced by the growing thin disk for the last
$\sim 9$ Gyr. Therefore, 
 previous simulations
which did not consider the formation of the thin disk subsequent to the formation
of the thick disk
are not ideally suited to a  comparison with the observed ones. 
Although the Galactic  bar (or the ``barred bulge'')
could have significantly influenced the dynamical evolution
of the thick disk, such dynamical influences have not been investigated
in previous simulations.  Furthermore, chemical properties of the thick disk formed
by ancient minor merger events
are yet to be discussed in the literature.

The purpose of this paper is thus to investigate chemical and dynamical properties
of the Galactic thick disk formed from ancient minor merger events based on new
chemical and dynamical models for the formation of the thick disk.
In the present minor merger scenario, the thick disk was formed from an
accretion of a dwarf galaxy on the FGTD about $\sim 9$ Gyr ago and
therefore has long been dynamically influenced by the growing thin disk since 
its formation. 
Although previous numerical simulations investigated the dynamical evolution
of the Large Magellanic Cloud
composed of thin and thick disks (Bekki 2009)
and dynamical influences of a growing thin disk
on a pre-existing thick disk in a disk galaxy (e.g., Kazantzidis et al. 2009;
Villalobos et al. 2010),  they did not discuss how non-axisymmetric structures
such as bars and spirals in a {\it live}  thin disk influence the dynamical evolution
of a thick disk in a galaxy.
The present study therefore adopts a new two-component disk  model in which
a {\it live} thin disk can 
dynamically influence a thick disk through the growth of  non-axisymmetric structures
over  a long timescale.
We then use the model to discuss
structural and kinematical properties of the present Galactic thick disk surrounding
the thin disk.

We also construct a new self-consistent chemical evolution model
in which the early star formation history and 
chemical evolution of the thin disk is 
influenced  by (i) the remaining gas of the thick disk,
(ii) later gas accretion onto the thin disk, and (iii)
dynamical properties of the thick disk. 
Since the present chemical evolution model is the first one that is based
on the minor merger scenario,  the predicted properties of the model
enable us to discuss strength and weakness  of the minor merger scenario in 
explaining the observed properties of the thick disk
(e.g., Chen et al. 2000; Gratton et al. 2000;
Prochaska et  al. 2000;
Mashonkina \& Grhren 2001;
Tautvaisiein\. e et al. 2001;
Bensby et al. 2003, 2005;
Reddy et al. 2003; 2006;
Casagrande et al. 2011).
Based on our  chemical and dynamical models,
we particularly discuss  the radial metallicity gradient of the thick disk,
dynamical interaction between the thin and thick disks after the formation of the
thick disk, the metallicity distribution function,  and the [$\alpha$/Fe]-[Fe/H] relation
of the thick disk. The present model is novel,
because these thick disk properties have not been discussed in previous
theoretical studies  based on the minor merger scenario.

The structure  of the paper is as follows. In the next section,
we  describe our new two-component dynamical  models
of the Galactic disk and our chemical evolution models for the Galactic thin
and thick disks. 
In \S 3, we present the results of the present chemical and dynamical
models for the Galactic disks. 
In \S 4, we discuss a number of key  observational results in the context
of different formation  scenarios for the thick disk and
we summarize our  conclusions in \S 5.
Although our previous chemodynamical models of the formation of the Galaxy
discussed  physical correlations
between [Fe/H] and kinematic of the Galactic disk and stellar halo
(Bekki \& Chiba 2000, 2001), the present study, which is not
based on a fully self-consistent chemodynamical model, does not allow us to
do so.  Because of this limitation, we only briefly discuss the observed
correlations between chemical and kinematical properties of the thick disk
in the present paper,  and will extensively discuss these in our future studies
based on fully self-consistent chemodynamical simulations. 
Because we focus exclusively on the Galactic thick disk in the present paper,
we  will
discuss the origin of thick disks observed
in other disk galaxies (e.g., Yoachim \& Dalcanton 2008)
in the context of the minor merger scenario  in forthcoming papers.

\begin{deluxetable*}{cccccccccc}
\footnotesize
\tablecaption{Model parameters for N-body simulations.
\label{tbl-1}}
\tablewidth{0pt}
\tablehead{
\colhead{  Model } &
\colhead{  $M_{\rm d}$  \tablenotemark{a} } &
\colhead{  $f_{\rm dm}$  \tablenotemark{b} } &
\colhead{  $m_2$ \tablenotemark{c} }  &
\colhead{  $r_{\rm p}$  \tablenotemark{d} }  &
\colhead{  $e_{\rm p}$  \tablenotemark{e}}  &
\colhead{  $\Theta$  \tablenotemark{f} } &
\colhead{  $f_{\rm thick}$ \tablenotemark{g} }  &
\colhead{  $M_{\rm d,n}$ \tablenotemark{h}  } & 
\colhead{  $a_{\rm d,n}$ \tablenotemark{i}  } }
\startdata
M1 & 4.0 & 157 & 0.2 & 1.0 & 0.5 & 30 & 0.12 & 4.0 & 0.2 \\
M2 & 4.0 & 157 & 0.05 & 1.0 & 0.5 & 30 & 0.105 & 4.0 & 0.2 \\
M3 & 4.0 & 157 & 0.1 & 1.0 & 0.5 & 30 & 0.11 & 4.0 & 0.2\\
M4 & 4.0 &157 & 0.3 & 1.0 & 0.5 & 30 & 0.13  & 4.0 & 0.2 \\
M5 & 4.0 & 157 & 0.2 & 0.5 & 0.8 & 30 & 0.12  & 4.0& 0.2 \\
M6 & 4.0 & 157 & 0.1 & 1.0 & 0.5 & 60 & 0.12  & 4.0 & 0.2\\
M7 & 4.0 &157 & 0.2 & 1.0 & 0.5 & 30 & 0.12 & 4.0 & 0.14 \\
M8 & 8.0 &79 & 0.2 & 1.0 & 0.5 & 30 & 0.24 & 4.0 & 0.2 \\
M9 & 4.0 & 99 & 0.2 & 1.0 & 0.5 & 30 & 0.12  & 4.0 & 0.2 \\
M10 & 4.0 & 99 & 0.05 & 1.0 & 0.5 & 30 & 0.105  & 4.0 & 0.2\\
M11  & 4.0 & 99 & 0.2 & 1.0 & 0.5 & 30 & 0.12  &  4.0 & 0.14\\
M12  & 4.0 & 157 & 0.2 & 1.0 & 0.5 & 30 & 0.24  &  2.0 & 0.2\\
M13  & 4.0 & 157 & 0.2 & 1.0 & 0.5 & 30 & 0.40  &  1.2 & 0.2\\
M14  & 4.0 & 157 & 0.05 & 1.0 & 0.8 & 30 & 0.105  &  4.0 & 0.2\\
M15  & 4.0 & 157 & 0.1 & 1.0 & 0.5 & 0 & 0.11  &  4.0 & 0.2\\
M16  & 4.0 & 157 & 0.2 & 1.0 & 0.5 & 0 & 0.12  &  4.0 & 0.2\\
\enddata
\tablenotetext{a}{The total mass of the stellar disk in the FGTD in units 
of $10^{9} M_{\odot}$.}
\tablenotetext{b}{The mass ratio of the dark matter halo to the stellar disk
in the FGTD.}
\tablenotetext{c}{The mass ratio of the dwarf disk to the FGTD in a minor merger.} 
\tablenotetext{d}{The pericenter distance of the orbit in a minor merger model
in units of $R_{\rm d}$, where $R_{\rm d}$ is the disk size of the FGTD.} 
\tablenotetext{e}{The orbital pericenter distance in a minor merger model.}
\tablenotetext{f}{The inclination angle between the orbital plane  of the dwarf disk
and the $x$-$y$ plane (corresponding to the disk plane of the FGTD).} 
\tablenotetext{g}{The mass ratio of the thick disk to the thin disk in a two-component
disk model.}
\tablenotetext{h}{The total mass of the stellar disk in the thin disk  in units 
of $10^{10} M_{\odot}$.}
\tablenotetext{i}{The scale-length of the  thin disk  in units 
of the disk size ($R_{\rm d, n}$).}
\end{deluxetable*}

\section{Models}

We discuss chemical and dynamical properties of the Galactic thick disk
by using both N-body simulations
and one-zone chemical evolution models.
We first describe the models for the ancient minor mergers, the growth of the
thin disk, and the dynamical evolution of the stellar disk composed of
thin and thick disks (\S 2.1). These models are used for 
deriving  radial metallicity gradients of the thick disk formed by minor merging,
and determining  structural and kinematical properties of the present thick disk.
We then describe one-zone chemical evolution models that are used
for deriving the MDFs 
and correlations between different chemical abundances of the thin
and thick disks (\S 2.2).
Dynamical properties of the stellar remnants of minor mergers
and orbital eccentricity distributions of thick disk stars
formed from minor merging  have been
extensively discussed in a number of recent papers
(e.g., Villalobos \& Helmi 2008; Di Matteo et al. 2010; Qu et al. 2011)
and we confirm that their main results can be held for the present minor merger
models.
We therefore  focus on the  new
results that have not been discussed elsewhere.

\subsection{Dynamical models}

\subsubsection{FGTD}

The FGTD is modeled as
a bulge-less disk galaxy with total mass $M_{\rm d}$
and size $R_{\rm d}$ embedded in a massive dark matter halo.
The total  mass and the virial radius  of the dark matter halo  of the FGTD
are denoted as $M_{\rm dm}$ and $r_{\rm vir}$, respectively.
We adopt an NFW halo density distribution (Navarro, Frenk \& White 1996)
suggested from CDM simulations:
\begin{equation}
{\rho}(r)=\frac{\rho_{0}}{(r/r_{\rm s})(1+r/r_{\rm s})^2},
\end{equation}
where  $r$, $\rho_{0}$, and $r_{\rm s}$ are
the spherical radius,  the characteristic  density of a dark halo,  and the
scale
length of the halo, respectively.
We adopted $c=12$ ($=r_{\rm vir}/r_{\rm s}$)
and $r_{\rm vir}=12R_{\rm d}$ for the dark matter halo,
and the mass ratio of halo to disk ($f_{\rm dm}$)
is regarded as a free parameter that can control the mass-ratio
of the simulated thick disk to the thin disk.
In the present scenario,  minor merging occurred when the FGTD was still rapidly
growing and thus a much less massive disk embedded in a massive dark matter halo.
We therefore consider that $f_{\rm dm}$ should be rather large
and investigate models with $f_{\rm dm}$ ranging from 79 to 157.
We mainly describe the results of the models with $f_{\rm dm}=157$ in which
the final stellar disk has a  
thick disk component with the mass  $\sim 10$\% of
its thin disk component
and a maximum circular velocity ($V_{\rm c}$)  $\sim 240$ km s$^{-1}$, which  is
reasonable for the Galaxy (e.g., Binney \& Tremaine 2007).

The stellar component  of the FGTD is assumed to have an exponential
profile with  radial and vertical scale lengths of $0.2R_{\rm d}$
and $0.02R_{\rm d}$,
respectively.
In addition to the rotational velocity made by the gravitational
field of the disk and halo components, the initial radial and azimuthal velocity
dispersion are given to the disk component according
to the epicyclic theory with Toomre's parameter (Binney \& Tremaine 1987) $Q$ = 1.5.
The vertical velocity dispersion at a given radius 
is set to be 0.5 times as large as
the radial velocity dispersion at that point. 
The initial disk plane of the FGTD is set to be the $x$-$y$ plane
in the present study (i.e., the $z$-axis is the polar direction of the disk).
Figure  1 shows the radial profile of the circular  velocity  
$V_{\rm c}$ for the FGTD in a  model with 
$M_{\rm d}=4 \times 10^{9} M_{\odot}$ and $R_{\rm d}=17.5$ kpc.
It is clear from this figure that the original thin disk has $V_{\rm c}$ that
is significantly smaller than the present one in
the Galaxy.

Recent photometric and spectroscopic observations of the Galactic
stellar and gaseous components
have demonstrated that the present thin disk has metallicities that depend on
location within the disk, i.e., metallicity gradients
(e.g., Friel 1995).
In the present minor merger scenario,  stars within  the present
Galactic bulge had been already formed before the minor merger occurred
and they initially consisted of the inner part of the FGTD.
The inner part of the thick disk 
became the part of the Galactic bulge in response  to the bar formation in the
inner part of the two-component Galactic disk.
We thus need to assume that the inner disk of the FGTD has a  metallicity gradient
that is different from that of the outer part and similar to that of the
bulge.  In the present study, we consider that the metallicity gradient
is different between the inner ($R<R_{\rm th}$)  and outer ($R \ge R_{\rm th}$)
regions of the FGTD, where 
$R<R_{\rm th}$ is set to be 2 kpc,  corresponding to the size of the bulge.

We allocate metallicity to each disk star in the outer disk
($R \ge R_{\rm th}$)  according to its initial position:
at $r$ = $R$,
where $r$ ($R$) is the projected distance (in units of kpc)
from the center of the disk, the metallicity of the star is given as:
\begin{equation}
{\rm [m/H]}_{\rm r=R} = {\rm [m/H]}_{\rm d, r=0} + {\alpha}_{\rm d} \times {\rm R}. \;
\end{equation}
We consider  that (i) the slope ${\alpha}_{\rm d}$ is a free parameter and
(ii) ${\rm [m/H]}_{\rm d, r=0}$ is determined such that the metallicity
at the solar radius ($R=R_{\odot}$ corresponding to 8.5 kpc) is $-0.7$
for the adopted ${\alpha}_{\rm d}$.
We adopt the observed  value  of ${\alpha}_{\rm d} \sim -0.04$ 
(e.g., Andrievsky et al. 2004). 
Guided by  the observational results on the radial metallicity gradient
of the Galactic bulge 
(e.g., McWilliam \& Rich 1994; Wyse et al. 1997; Frogel et al. 2000),
we assign the metallicity of a star at $r$ = $R$ 
in the inner disk ($R<R_{\rm th}$),
where $r$ ($R$) is the projected distance (in units of kpc) of the star
from the center of the disk, to be:
\begin{equation}
{\rm [m/H]}_{\rm b, r=R} = {\rm [m/H]}_{\rm b, r=0} + {\alpha}_{\rm b} \times {\rm R}. \;
\end{equation} 
We allow the slope of the metallicity gradient to be  a free parameter and
investigate different models with different ${\alpha}_{\rm b}$.
The central metallicity of the disk is determined such that 
the metallicity at $R=R_{\odot}$ is consistent with the value  derived from
the  equation (2) for the outer disk.
If we adopt the metallicity gradient from Frogel et al. (2000) in which
the slope ${\alpha}_{\rm b} = -0.4$, 
then ${\rm [m/H]}_{\rm b, r=0} = 0.36$ for ${\alpha}_{\rm d} = -0.04$
and ${\rm [m/H]}_{\rm d, r=0} = -0.36$.

The dwarf disk is also assumed to have a metallicity gradient, though
it does not have an inner disk component that  becomes a bulge. 
Therefore, the radial metallicity gradient for the entire disk region
is described by the following single relationship:
\begin{equation}
{\rm [m/H]}_{\rm r=R} = {\rm [m/H]}_{\rm dw, r=0} + {\alpha}_{\rm dw} \times {\rm R}. \;
\end{equation}
The slope of the metallicity gradient ${\alpha}_{\rm dw}$ is assumed to be 
a free parameter that can control the final metallicity gradients of merger remnants
(i.e., thick disks).
Since there is an observed relation between luminosity ($L$) and metallicity ($Z$)
for galaxies ($Z$ $\propto$ $L^{0.4}$; Mould 1984),
we allocate a smaller  metallicity of ${\rm [m/H]}_{\rm dw, r=0}$
for the dwarf disk according to the mass of the stellar disk
in the dwarf ($M_{\rm d,dw}$).
The dwarfs disks in the present study are assumed to have total masses
being $5-20$\% of the FGTD 
and thus cannot be similar to low-mass dwarfs with lower [$\alpha$/Fe]
observed in the local group: they are not literally dwarfs. The adopted massive dwarf disks 
can experience rapid star formation in their  early histories
(thus have higher [$\alpha$/Fe])  and correspond to 
the Large Magellanic Cloud which shows higher [$\alpha$/Fe] in its old stellar populations.
Thus the simulated thick disks in the present study,
which can have stars from disrupted dwarf disks,  can show higher [$\alpha$/Fe],
which is consistent with recent observations by Ruchti et al. (2011) and Lee et al. (2011).

In order to show more clearly how minor galaxy merging can influence
radial metallicity gradients of merger remnants,
we allocate metallicities to stars at each radius according to their
positions $R$ alone. We do not introduce initial vertical metallicity gradients
in the FGTD in the present study, because we need to assume a few additional unknown
parameters, which can make the interpretation of the simulation results much less
straightforward. It would be  possible for different stars at the same $R$
in the FGTD to have different metallicities owing to chemical and dynamical evolution,
but we do not consider this possibility.
Ignoring such a  dispersion in 
the initial metallicity at each radius
allow us  to avoid introducing additional free parameters for clarity.
The final mean metallicity in the central region ($R<2$ kpc) 
corresponding to the Galactic bulge
region in each merger remnant 
can be as low as $-0.3$,  owing to radial mixing of stellar populations
in the present study. The central metallicity can be higher if we adopt
higher metallicities in the inner regions of the FGTD. Since the purpose of this paper
is not to discuss the origin of the bulge, we do not consider its chemical properties.

The total numbers of particles used for the dark matter halo 
($N_{\rm dm}$) and  the stellar disk ($N_{\rm disk}$) 
of  the FGTD
in each  simulation are 800000 and 100000, respectively.
The total number of particles used for the dark matter halo 
and  the stellar disk in a dwarf are
$m_2 N_{\rm dm}$ and $m_2 N_{\rm disk}$, respectively,
where $m_2$ is the mass ratio of the dwarf to the FGTD.
As demonstrated by Waker et al. (1996),
the total particle number of more than $5 \times 10^5$ is enough
to properly investigate the formation processes of thick disks from
thin ones through minor galaxy merging.
Therefore, the present simulations with $N \sim 10^6$ enable us to
derive physical properties of galactic thick disks in a convincing way.
In all of the present models, 
the adopted  gravitational softening
length is fixed at $0.014R_{\rm d}$, which corresponds to 252pc for
$R_{\rm d}=17.5$ kpc.

\begin{figure}
\plotone{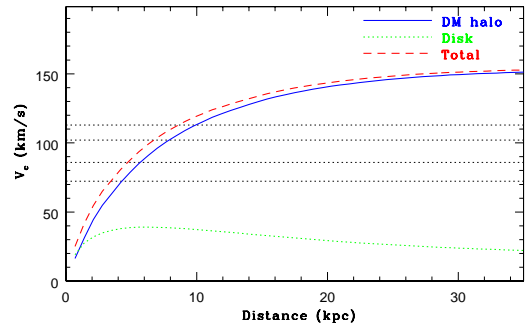}
\figcaption{
The radial profile of the 
circular velocity  ($V_{\rm c}$; red dashed) 
for  the FGTD with $M_{\rm d}=4 \times 10^9 M_{\odot}$,
$R_{\rm d}=17.5$ kpc, and $f_{\rm dm}=157$: contributions from the  dark matter
halo and the stellar disk are shown by blue solid and  green dotted lines,
respectively. The maximum circular velocity of a dwarf disk that merges with
the FGTD is shown by a dotted line for $m_2=0.05$, 0.1, 0.2, and 0.3, where $m_2$
is the mass ratio of the dwarf to the FGTD.
\label{fig-1}}
\end{figure}

\begin{figure}
\plotone{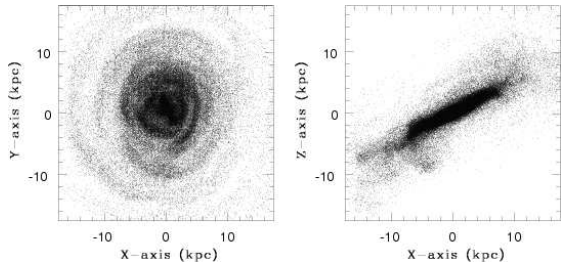}
\figcaption{
The final stellar distribution projected onto the $x$-$y$ plane (left)
and the $x$-$z$ plane (right)
for the minor merger model with $m_2=0.05$, $r_{\rm p}=R_{\rm d}$,
$e_{\rm p}=0.5$, and $\Theta=30$. 
\label{fig-2}}
\end{figure}

\begin{figure}
\plotone{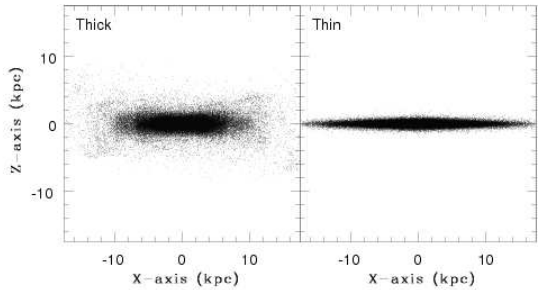}
\figcaption{
The initial  stellar distribution projected onto the $x$-$z$ plane 
for the thick disk component (left) and thin one (right)
in the two component disk model
constructed from the remnant of the minor merger model 
with $m_2=0.05$, $r_{\rm p}=R_{\rm d}$,
$e_{\rm p}=0.5$, and $\Theta=30$. 
\label{fig-3}}
\end{figure}

\subsubsection{Minor mergers}

We consider minor galaxy mergers with  mass ratios ($m_{\rm 2}$)
of merger progenitor
galaxies ranging from 0.05 to 0.3,  so that we can investigate the physical properties
of thick disks. The FGTD and the satellite galaxy that finally  merges with the FGTD  
are assumed to have self-similar structures and kinematics in dark matter halos
and stellar disks: they both have NFW halos with 
$R_{\rm vir}/R_{\rm d}=12$ and exponential disks with scale lengths and
vertical scale heights of $0.2R_{\rm d}$ and $0.02R_{\rm d}$, respectively.
We investigate only galaxy mergers between bulge-less stellar disks in the present
study, though there can  be some differences in dynamical properties
of merger remnants between disk-disk merger models and disk-spheroidal ones
(e.g., Villalobos et al. 2008). Since our main focus is not the detailed
properties of minor merger remnants, we do not discuss the results of disk-spheroidal
merger models in the present study.
The disk size of the dwarf ($R_{\rm d, dw}$) is assumed to depend on 
the mass of the disk  ($M_{\rm d, dw}$) such as
$R_{\rm d, dw} \propto M_{\rm d,dw}^{0.5}$, which corresponds to
Freeman's law (Freeman 1970). Therefore, $m_{2}$ can determine
the size of the stellar disk for the dwarf and thus the structural and 
kinematical  properties of the dark matter halo and the stellar disk.

In the present simulation of a minor merger, the orbit of the
two disks is set to be initially in the $x$-$y$ plane,  and the
orbital plane of the satellite disk is assumed to be inclined
by ${\Theta}$ degrees with respect to the orbital plane.
The initial distance between the two spirals ($r_{\rm i}$)
is set to be either $3R_{\rm d}$ or $4R_{\rm d}$ in the present study. 
The pericenter distance ($r_{\rm p}$)  and the orbital
eccentricity ($e_{\rm p}$) are assumed to be free parameters
which control the orbital angular momentum and the energy of the merging
galaxies. For most merger models, $r_{\rm p}$ and
$e_{\rm p}$ are set to be 1.0 (in our units) and 1.0,
respectively.
The spin of the satellite dwarf galaxy 
is specified by two angles $\theta$ and
$\phi$ (in units of degrees).
Here, $\theta$ is the angle between
the $z$-axis and the vector of the angular momentum of the disk,
and $\phi$ is the azimuthal angle measured from the  $x$ axis to
the projection of the angular momentum vector of the disk onto
the $x$-$y$ plane. We show the results of the models with
$\theta=45$ and $\phi=30$ in the present study.
We investigate the dynamical evolution of a minor merger 
for $[32-40] \times t_{\rm dyn}$
in each  model (depending on $m_2$ and orbital configurations), where
$t_{\rm dyn}$ is the dynamical time scale of the FGTD.
We adopt  $t_{\rm dyn}= 1.7 \times 10^8$ yr
in the present study.

Figure 2 shows an example of the final stellar distributions
in the stellar remnants of minor galaxy mergers.
The initial thin disk in this merger model has
$m_2=0.05$, $\Theta=30$, $r_{\rm p}=R_{\rm d}$, and $e_{\rm p}=0.5$
has $M_{\rm d}=4 \times 10^9 M_{\odot}$, $R_{\rm d}=17.5$ kpc,
and $f_{\rm dm}=157$. The thin disk can be dynamically heated up
by the sinking dwarf disk so that the disk can be transformed into a thick disk
with outer disturbed stellar substructures. The vertical stellar velocity dispersion
and the rotational velocity in the thick disk
are $\sim 30$ km s$^{-1}$ and $\sim 120$ km s$^{-1}$ 
at $R=R_{\odot}$ (=8.5 kpc), respectively.
 The stellar velocity dispersions and rotational velocity
of the thick disk can increase significantly during the growth of a thin disk
surrounded by the thick disk. As described later,
the final rotational velocity  of the thick disk 
(formed in  this  minor merger model) after the growth of the thin disk
is not altogether  consistent with observations. In the present study, merger remnants
with larger $m_2$ ($\sim 0.2$)  can be regarded as better models for the thick disk formation.

\subsubsection{Slow growth of thin disks for  two-component disk models}

A thin stellar disk is placed within the thick disk 
so that the  dynamical response of the thick disk to the
slowly growing thin disk  can be investigated.  The thick disk formed 
in a minor merger model is inclined significantly with respect to
the $x$-$y$ plane with the inclination angle depending on $\Theta$. 
Accordingly a coordinate transformation is  performed for the thick disk,
so that the final disk plane can be again
coincident with  the $x$-$y$ plane
and thus with the disk plane of the thin disk. 
Then the stellar particles that finally consist of the new thin disk
are slowly added to the thick disk 
until the thin disk finally has a mass of $M_{\rm d, n}$ and a size of
$R_{\rm d, n}$.  This addition of stellar particles can mimic the slow
growth of the thin disk by gas accretion from the Galactic halo region
and the subsequent star formation from the gas.
The accretion/formation  timescale of the thin disk
is set to be $20t_{\rm dyn}$ for all models in 
the present study. The values of $M_{\rm d, n}$ and $R_{\rm d, n}$
are fixed at $4 \times 10^{10} M_{\odot}$ and $17.5$ kpc, respectively, for most models
so that physical
properties of  the simulated thin disk
can be consistent with the observed  ones. 
During the growth of the thin disk,  the thin disk particles keep
their original locations and velocities,  whereas the thick disk particles
are allowed to dynamically respond to the changing gravitational potential
due to the growth of the thin disk.

The thin disk is assumed to be an exponential disk with
the radial scale length being either $0.143R_{\rm d,n}$ or $0.2R_{\rm d,n}$
and  the vertical scale height being 
$0.02R_{\rm d, n}$. 
Stellar particles with the total particle number $n_{\rm acc}$ 
are randomly allocated their initial
locations at each time step
during the growth of the thin disk. We assume that the disk growth rate
is constant,  $n_{\rm acc}$ at each time step is $N_{\rm d, n}/N_{\rm step}$
in the present study,
where $N_{\rm d, n}$ is the total number of stellar particles used in the
thin disk and $N_{\rm step}$ is the total number of time steps in a numerical
simulation for the growth of the thin disk.
After the completion of the thin disk formation, we then allocate the three
dimensional (3D) velocities to each thin disk stellar particle according to
the final mass profiles of the dark matter, thick disk, and thin disk 
in the same way that we did for the FGTD.
In addition to the rotational velocity made by the gravitational
field of halo and disk components, the initial radial and azimuthal velocity
dispersion are given to the disk component according
to the epicyclic theory with Toomre's parameter $Q$ = 1.5.
The total number of stellar particles in the thin disk is set to be 100000 for all
models in the present study. Thus the two-component disk consists of
 collisionless particle with the same softening length
for thin and thick disks  and the total particle number of
$900000\times(1+m_2)+100000$
in the simulation.

Figure 3 shows the final stellar distributions of the thick and thin disk components
in the two-component disk constructed from the remnant of the minor  merger 
shown in Figure 2. The mass ratio of the thick disk 
to the thin one ($f_{\rm thick}$)
is 0.105 and $a_{\rm d, n}=0.2R_{\rm d, n}$
in this model.
Clearly, the thick disk has a larger scale-height and still 
shows a diffuse flattened halo-like structure ($R>10$ kpc). The original thick
disk is dynamically compressed and stellar substructures formed in minor merging
cannot be clearly seen in this two-component disk. The vertical velocity dispersion
and rotational velocity of the thick disk
finally becomes 39 km s$^{-1}$ and 187 km s$^{-1}$ at $R=R_{\odot}$, respectively,
both of which are significantly
higher than the original values in the thick disk before the slow growth of the 
thin disk. The final vertical velocity of the thick disk is larger by a factor of 2.8
 than that of the thin disk in this two-component disk model.
Although we mainly investigate dynamical evolution of two-component disk models 
with  $a_{\rm d, n}=0.2R_{\rm d, n}$, we compare the results between models
with $a_{\rm d, n}=0.143R_{\rm d}$ and $0.2R_{\rm d, n}$.

\subsubsection{A parameter study}

We investigate the dynamical evolution of 
the thin and thick disks for $40t_{\rm dyn}$
by using the two-component stellar disks constructed
above. We particularly investigate rotational velocities
($V_{\phi}$), radial (${\sigma}_{\rm r}$),
azimuthal (${\sigma}_{\phi}$), and vertical (${\sigma}_{\rm z}$) velocity dispersions
in the thin and thick stellar disks when the central
stellar bar (=boxy bulge) is formed for each model.
Although we have run models with different model parameters,
we mainly describe  the results of the ``standard  model'' 
with $M_{\rm d}=4 \times 10^9 M_{\odot}$,
$R_{\rm d}=17.5$ kpc,
$f_{\rm dm}=157$,
$m_2=0.2$,
$r_{\rm p}=R_{\rm d}$,
$e_{\rm p}=0.5$,
$\Theta=30$,
$M_{\rm d, n}=4 \times 10^{10} M_{\odot}$,
and $R_{\rm d,n}=17.5$ kpc.
Figure 4 shows the radial profiles of $V_{\rm c}$ and $V_{\phi}$ in 
the thin and the thick disk for the two-component disk model.
The standard model 
shows larger differences in $V_{\phi}$ ($\sim 30$ km s$^{-1}$),
which is  more consistent with observations (e.g., Soubiran et al. 2003)
in comparison
with other models.
The mass-ratio of the thick disk 
to the thin one ($f_{\rm thick}$)
is 0.12, which is consistent with the observed range of the mass-ratio.

We consider that the models with higher $m_2$ ($\sim 0.2$) can be better ones
for the thick disk formation in comparison with those with lower $m_2$ ($\sim 0.05$),
mainly because  differences in $V_{\phi}$ in the simulated thin and thick disks
in the models with larger $m_2$ 
are more similar to the observed one (e.g., Soubiran et al. 2003).
The two-component disk model shown in Figure 3 shows a rather small difference
in $V_{\phi}$ at $R=R_{\odot}$ (at most $\sim 10$ km s$^{-1}$) so that the model cannot
be the best model. Two-component disk models 
constructed from stellar remnants
of more energetic mergers  with $m_2=0.05$, $r_{\rm p}=0.5R_{\rm d}$,
and large $e_{\rm p}$ (e.g., 0.8) however show
larger differences in  $V_{\phi}$ between the thin and thick disks.
We investigate the following four 
kinematical properties at $R=R_{\odot}$
in each model:
(i) the circular velocity $V_{\rm c}$,
(ii) the difference in $V_{\phi}$ between the thin and thick disks, 
(iii) ${\sigma}_{\rm z}$,
and (iv) ${\sigma}_{\rm r}/{\sigma}_{\rm z}$,
${\sigma}_{\phi}/{\sigma}_{\rm z}$,
and ${\sigma}_{\rm r}/{\sigma}_{\phi}$. 
Since the entire structure and kinematics of the thick disk are currently unknown,
we compare  the above kinematical properties at $R=R_{\odot}$ 
with the corresponding observational ones (e.g., Dinescu et al. 2011)
for the diagnosis of
the better models.

Most of the present models  show
higher ${\sigma}_{\rm z}$ ($40-50$ km s$^{-1}$) at $R=R_{\odot}$ in the thick
disk component, which is  slightly larger than the observed value.
The present minor merger simulations are purely collisionless ones
so that they can overpredict stellar velocity dispersions of merger
remnants (=thick disks)   owing to the lack
of gaseous dissipation and cooling. 
We therefore consider that  models with ${\sigma}_{\rm z} \sim 50$ km s$^{-1}$
that is appreciably larger than the observed one ($40 \pm 10$ km s$^{-1}$;
Freeman 1986; $39 \pm 4$ km s$^{-1}$ Soubiran et al. 2003)
can be regarded as  reasonable. 
The present minor models are less realistic in some points (e.g., non-inclusion
of gas dynamical and star formation), and we have explored
a limited range of model parameters. Therefore, it should be stressed  that
the the present standard model does  not explain all of the observed properties
of the thick disk in a fully self-consistent manner.

The Table 1 summarizes the parameters values for the representative models
investigated in the present study. The standard model is labeled  as M1. 
For each of these models, we investigate the radial metallicity gradient
and the MDF of the simulated thick disk and their dependences on the model
parameters for the initial metallicity gradient of the FGTD (e.g., ${\alpha}_{\rm d}$).
Numerical computations
were carried out both  on
(i) the latest version of GRAPE
(GRavity PipE, GRAPE-DR) -- which is the special-purpose
computer for gravitational dynamics (Sugimoto et al. 1990)
and (ii)  one IBM system iDataPlex
with two GPU cards (NVIDIA Tesla M2050) and the CUDA G5/G6 software package 
installed for calculations of gravitational dynamics
at University of Western Australia.
It took about 9 CPU hours, 6 CPU hours,
and 14 CPU hours  for one GRAPE-DR machine  to simulate 
one minor merger model,   slow growth of a thin disk surrounded by a thick disk,
and dynamical interaction between the thin and thick disks, respectively
(thus about 29 CPU hours for $\sim 92t_{\rm dyn}$ dynamical evolution 
of  one sequential dynamical simulation).
The time $T$ represents the elapsed time for each  simulation of 
the two-component disk evolution.

\begin{figure}
\plotone{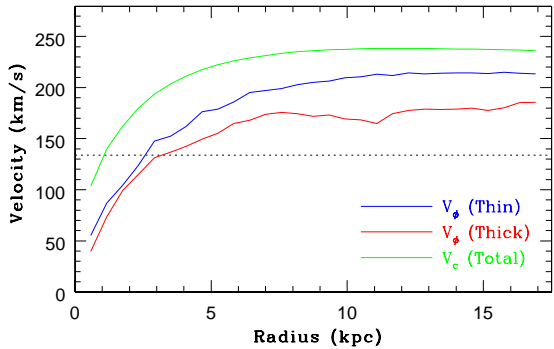}
\figcaption{
The radial profiles of the rotational velocity of the thin disk (blue),
that of the thick disk (red), and
the circular velocity (green) in the two-component disk for the standard model.
\label{fig-4}}
\end{figure}

\begin{figure}
\plotone{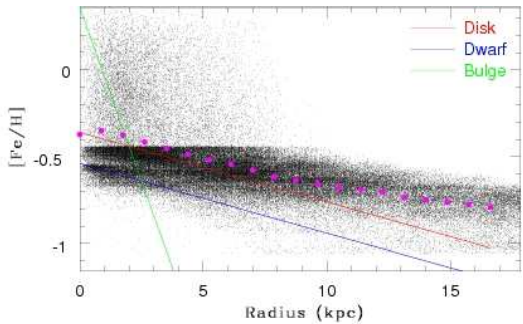}
\figcaption{
The radial distribution of stellar metallicities in the thick disk formed
in the minor merger model
with $m_2=0.2$, $r_{\rm p}=R_{\rm d}$,
$e_{\rm p}=0.5$, and $\Theta=30$, which is used for constructing the two-component
disk in the standard model. Each small black dot represent the location of a star
in the thick disk. The red, blue, and green line represent the initial metallicity
gradients for the stellar disk of the FGTD, the dwarf, and the bulge component 
of the FGTD, respectively. Each  magenta filled circle represents the
mean [Fe/H] for each radial bin. 
\label{fig-5}}
\end{figure}

\subsection{Chemical evolution models}

Since the present one-zone chemical evolution models for 
the Galaxy are  essentially the same
as those adopted in our previous paper (Tsujimoto et al. 2010),
we here briefly describe the models.
The basic picture is that the Galactic thin and thick disks
were  formed by gas infall from 
outside the disk region. For the infall rate, 
we apply an exponential form with a timescale $\tau_{\rm in}$. Taking into account 
the relatively rapid formation of the thick disk and the presence of 
the G-dwarf problem for the thin disk, we assume a rather short timescale of 
$\tau_{\rm in}$=0.3 Gyr for the thick disk, and a much longer timescale of $\tau_{\rm in}$=4 Gyr for the thin disk. The metallicity $Z_{\rm in}$ of infalling gas
is set up as follows. The MDF of the thick disk shows
a sharp increase from [Fe/H]$\sim -1.3$
to the peak located at [Fe/H]$\sim -0.7$. This feature was
 first reported by Wyse \& Gilmore (1995)
with a small number of complete samples, 
and the location of its peak is confirmed by the huge SDSS database of  
stars within the thick disk  (Allende et  al.~2006). 
These observational results suggest that  material for the proto-thick disk 
was pre-enriched up to [Fe/H]$\sim -1.3$. 
The mechanism of pre-enrichment may be attributable
to  wind enrichment triggered by an initial starburst 
in the Galactic bulge (Tsujimoto et al.~2010;  Tsujimoto 2011). 
Accordingly, [Fe/H]=$-1.3$ is assumed for the infalling gas of the thick disk.

On the other hand, $Z_{\rm in}$ of the thin disk
(i.e., metallicity of the gas accreted from the Galactic halo)
is determined by an implication from 
the cosmic evolution of damped Ly$\alpha$ systems (Wolfe et al.~2005).  
Its metallicity at the epoch of  thin  disk formation, 
i.e., $\sim 9$ Gyr ago, is  assumed to be fixed around [Fe/H]=$-1.5$ 
while  the thin disk is forming.
We also assume that the [$\alpha$/Fe] ratio in the gas accreted onto 
the thin disk is fixed at a high value (e.g., [Mg/Fe]=+0.4) 
as expected from chemical enrichment dominated by SNe II. 
As shown later in \S 3.3,
the present model can self-consistently reproduce some of the observed
chemical properties of the Galactic thin and thick disks
for the adopted values of $Z_{\rm in}$ and the [$\alpha$/Fe] ratios for
gas accreted onto the disks.

The star formation rate (SFR) is assumed to be proportional to the gas fraction with a constant coefficient $\nu$ for the duration $\Delta_{\rm SF}$. The higher value
of $\nu$=2 Gyr$^{-1}$ for $\Delta_{\rm SF}$= 1.5 Gyr is adopted for the thick disk, compared with $\nu$=0.7 Gyr$^{-1}$ and $\Delta_{\rm SF}$=12 Gyr for the thin disk. 
Here, the duration of thin disk formation is set so that 
the sum of timescales for two disk formation 
is broadly equivalent to the age of the Universe (=13.7 Gyr).
For the initial mass function (IMF), we assume a power-law mass spectrum with a slope of $-1.35$, which is combined with the nucleosynthesis yields of SNe Ia and II taken from Tsujimoto et al.~(1995). In our model, the products of SNe II are ejected with the
 short lifetime $\sim 10^{6-7}$ yr of massive stars depending on their masses, while those of SNe Ia are released with a delay of a considerably longer lifetime spanning
 over 0.5 - 3 Gyr (Yoshii et al.~1996). The fraction of stars that
 eventually produce SNe Ia for 3-8 $M_\odot$ is assumed to be 0.05.

In the present scenario  of  thick disk formation, 
star formation within  the thin disk can occur  after the termination of
star formation in the thick disk.  Accordingly, the present chemical evolution
models are different from those previously
adopted (e.g., Chiappini et al. 1997).
We consider the  three models for the star formation history of
the thin disk after the formation of the thick disk by minor merging described below. 
The first scenario is the ``continuous star formation model'' in which
the thin disk stars start to form
from the thick disk's remaining gas 
(corresponding to only $\sim 10$\% of the original gas)
mixed with the gas accreted onto the disk
{\it immediately after the thick disk formation}.
In this model, chemical abundances of the first stars in the thin disk
are similar to those of the most 
metal-rich stars  in the thick disk.

The second scenario  is the  ``temporal truncation model''
in which star formation in the thin disk cannot start until some amount of
gas is accreted onto the disk. Therefore, the first stars in the thin disk
will have  chemical abundances different from those of the final stars
in the thick disk. 
Here the timing to initiate the formation of thin disk stars 
is set at the time when the metallicity of gas is diluted to [Fe/H]=$-0.7$ 
by gas accretion from the halo region
(corresponding to about 1.5 Gyr after the formation of the  thick disk).
The total amount of the halo gas accreted onto the disk
before the commencement of  star formation in the thin disk
is  about five times larger than
that of the remaining gas of the thick disk.
We discuss a possible mechanism for the assumed temporal
truncation of star formation  later in Appendix A.
The third scenario  is the ``gas expulsion model''
in which 
the thick disk's remaining gas is 
expelled immediately after the formation of the thick disk 
and thus the thin disk is formed purely from gas accreted from the halo. 
Since our N-body simulations, which do not include star formation and gas accretion,
do not allow us to determine the most probable model among the above-mentioned three,
we investigate all  three scenarios and compare them with one another in \S 3.3.

\begin{figure}
\plotone{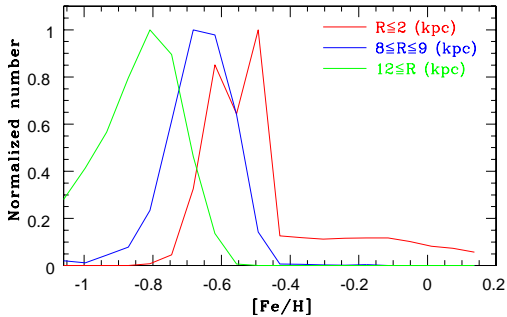}
\figcaption{
The MDFs for $R \le 2$kpc (bulge region; red), $8\le R \le 9$ kpc
(the solar neighborhood; blue), and $R \ge 12$ kpc (outer disk, green)
in the thick disk shown in Figure 5. The number distribution of stars 
is normalized by the maximum number of stars in 
the metallicity bins for each of the three regions. 
\label{fig-6}}
\end{figure}

\begin{figure}
\plotone{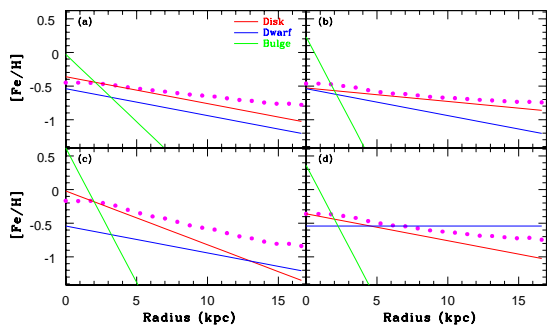}
\figcaption{
The same as Figure 5 but for four different models: (a) shallower metallicity
gradient in the bulge (${\alpha}_{\rm b}=-0.2$), (b) shallower metallicity
gradient in the disk ((${\alpha}_{\rm d}=-0.02$),
(c) steeper metallicity gradient in the disk (${\alpha}_{\rm d}=-0.08$),
and (d) no metallicity gradient in the dwarf (${\alpha}_{\rm dw}=0$).
\label{fig-7}}
\end{figure}

\begin{figure}
\plotone{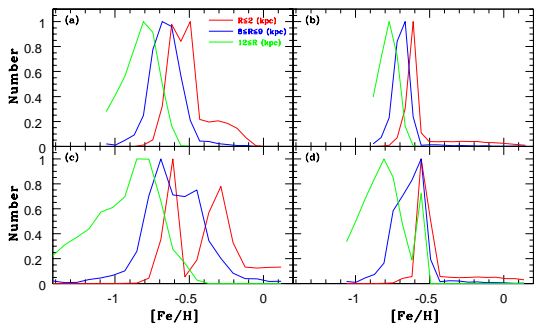}
\figcaption{
The same as Figure 6 but for four different models shown in Figure 7.
\label{fig-8}}
\end{figure}

\section{Results}

\subsection{Metallicity gradients and MDFs in thick disks}

Figure 5 shows the radial distribution of stellar metallicities in the thick
disk formed from minor  merging with $m_2=0.2$, $r_{\rm p}=R_{\rm d}$,
$e_{\rm p}=0.5$, and $\Theta=30$ (i.e., the standard model M1) for
${\alpha}_{\rm d}=-0.04$ ,
${\alpha}_{\rm b}=-0.4$,  and
${\alpha}_{\rm dw}=-0.04$.
During minor merging,
stellar populations with different metallicities can be radially mixed 
so that the original radial metallicity gradient changes significantly.
As shown clearly in Figure 5,
the initial steeper radial metallicity gradient of the FGTD can  become
significantly flattened after minor merging. More metal-rich stars initially
in the inner disk region  can be finally located at $R>5$ kpc
so that the final mean metallicities at radial bins with $R>5$ kpc can
slightly increase. On the other hand, the more metal-poor stars initially in the 
outer part of the FGTD can be located in the inner disk region, which lowers
the  mean metallicities for $R<2$ kpc.
The mean metallicities in radial bins with $R>10$ kpc can significantly
increase after minor merging, because the more metal-rich stars in the inner region
of the disk ($R<10$ kpc) can be transferred to the outer region.

The radial mixing of the  stellar population in minor merging can transfer metal-rich
stars with [Fe/H]$\sim -0.5$ $-$ $-0.3$ initially in the inner disk region 
into the solar neighborhood
($R=R_{\odot}$ corresponding to 8.5 kpc). 
This result implies that 
some fraction of metal-rich stars
currently at $R=R_{\odot}$ in the thick disk
can originate from the inner disk of the FGTD. It should be 
 stressed that the radial transfer of the inner metal-rich stars into
the solar neighborhood depends strongly on $m_2$ and  that the radial transfer
can be less effective in minor merging with smaller $m_2$. As a result of radial
mixing of stellar populations with different metallicities, the MDF of the thick
disk is different between different regions. Figure 6 shows that the MDF
in the solar neighborhood (8 kpc$\le R \le9$kpc) has a peak metallicity 
at [Fe/H]$\sim -0.7$ and a 
wide metallicity distribution.
The MDF of the thick disk at the bulge region ($R\le2$ kpc) shows a higher 
peak metallicity at [Fe/H]$\sim-0.5$ and a narrower distribution with a long tail
in the distribution at  higher [Fe/H] ($>0.1$).
The MDF at $R\ge12$ kpc shows a lower peak metallicity at [Fe/H]$\sim -0.8$ and
a larger fraction of metal-poor stars with [Fe/H]$<-0.8$.

The final radial metallicity gradient 
of the thick disk depends on the adopted
initial metallicity gradient of the FGTD.
We confirm that  flattening of the initial metallicity gradient of the FGTD
can be seen in almost all of the simulated thick disks in different
models for initial radial gradients of stellar metallicities,
though the degree of the flattening 
depends largely on $m_2$.
Figure 7 shows that the radial metallicity gradient of the thick
disk is not influenced  so much by that of the bugle, because only
a minor fraction of the bulge stars can be transferred to the disk region
during minor merging.  As shown in Figure 7,  the flattening of the
radial metallicity gradient can be more clearly seen in the model
with the steeper initial metallicity gradient of the FGTD.
There is no significant difference in the final metallicity gradient of
the thick disk between the standard model and the one with no initial
metallicity gradient in the dwarf disk, which means the transfer of 
of metal-poor stellar populations from  a dwarf disk to 
the thick disk is not
an important factor for  determining  the radial metallicity gradient
of the thick disk.
Figure 8 shows that the shapes of the  MDFs for the three different regions
of  the thick disk depend  on the adopted initial metallicity gradient
of the FGTD, which suggests that the MDFs can have some fossil information on
the original metallicity gradient of  the FGTD  before ancient  minor merging
transformed it into the thick disk.

\begin{figure}
\plotone{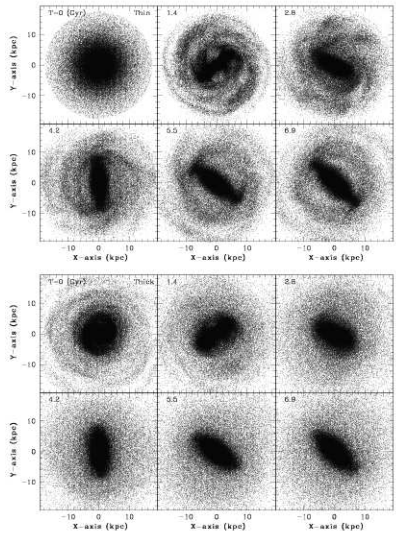}
\figcaption{
The time evolution of stellar distributions for the thin disk component
(upper six) and for the thick one (lower six) projected onto the $x$-$y$
plane in the two-component disk for the standard model. The time $T$ 
in units of Gyr
is shown in the upper left corner of each panel.
\label{fig-9}}
\end{figure}

\begin{figure}
\plotone{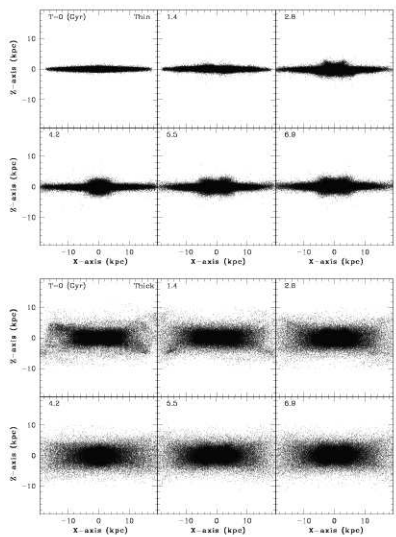}
\figcaption{
The same as Figure 9 but for the $x$-$z$ plane.
\label{fig-10}}
\end{figure}

\begin{figure}
\plotone{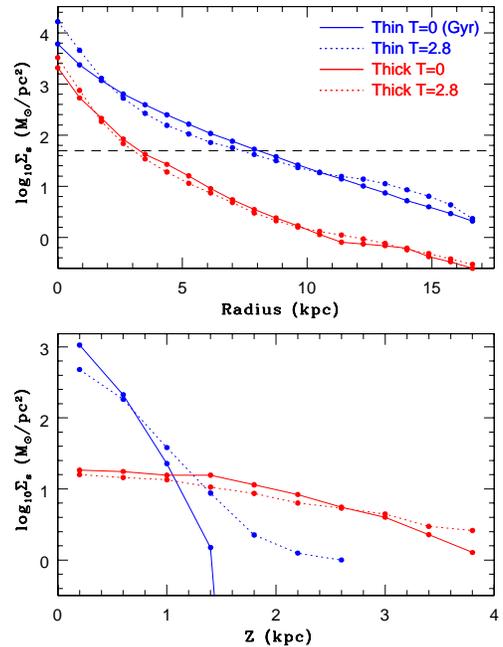}
\figcaption{
The radial (upper) and vertical (lower) stellar surface density (${\Sigma}_{\rm s}$)
profiles for the thin (blue) and thick (red) disks at $T=0$ Gyr (solid)
and $T=2.8$ Gyr (dotted) for the two-component disk in the standard model.
The black dashed line in the upper panel represents the observed ${\Sigma}_{\rm s}$
(49.6 $M_{\odot}$ pc$^{-2}$) at $R=R_{\odot}$.  
\label{fig-11}}
\end{figure}

\begin{figure}
\plotone{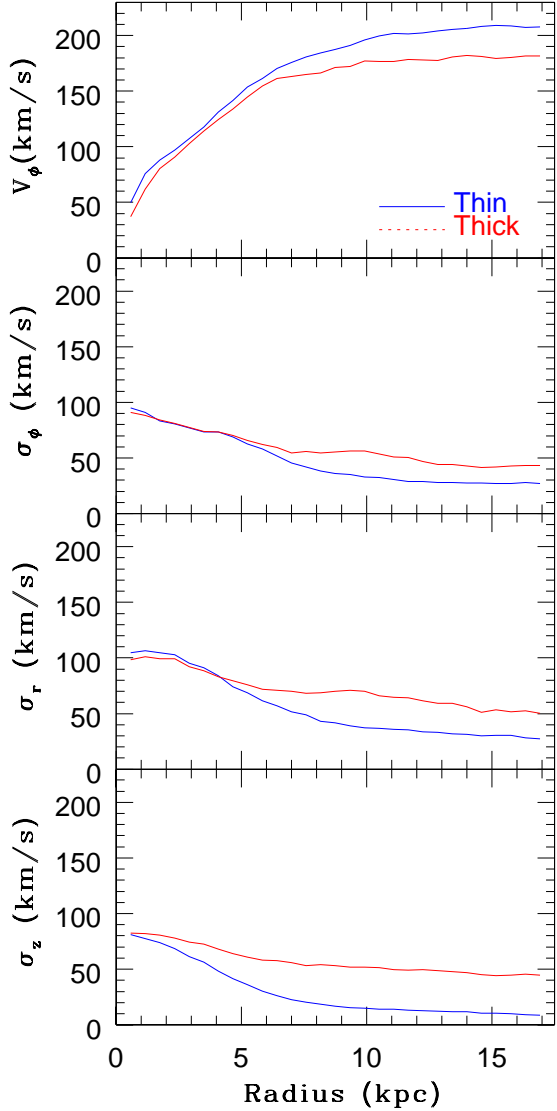}
\figcaption{
The radial profiles of $V_{\phi}$ (top), ${\sigma}_{\phi}$ (the second from the top),
${\sigma}_{\rm r}$ (the second from the the bottom),
and ${\sigma}_{\rm z}$ (bottom) for the thin disk (blue) and the thick disk (red).
\label{fig-12}}
\end{figure}

\begin{figure}
\plotone{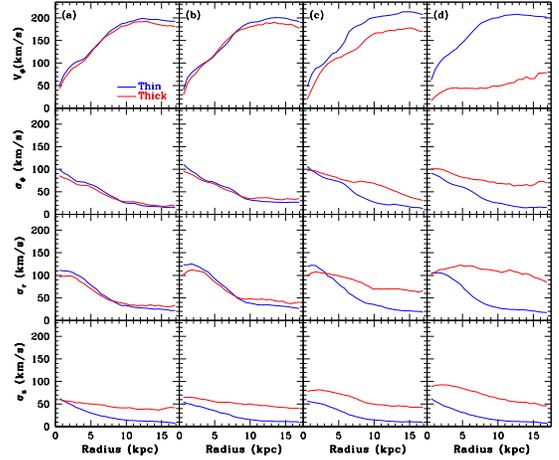}
\figcaption{
The same as Figure 12 but for four different models: (a) $m_2=0.05$
(model M2), (b) $m_2=0.1$ (M3),
(c) $m_2=0.3$ (M4), and (d) $m_2=0.2$, $r_{\rm p}=0.5R_{\rm d}$, and $e_{\rm p}=0.8$
(M5).
For the three models, (a), (b), and (c), the orbital parameters of minor 
mergers are the same as those adopted in the standard model. 
\label{fig-13}}
\end{figure}

\begin{figure}
\plotone{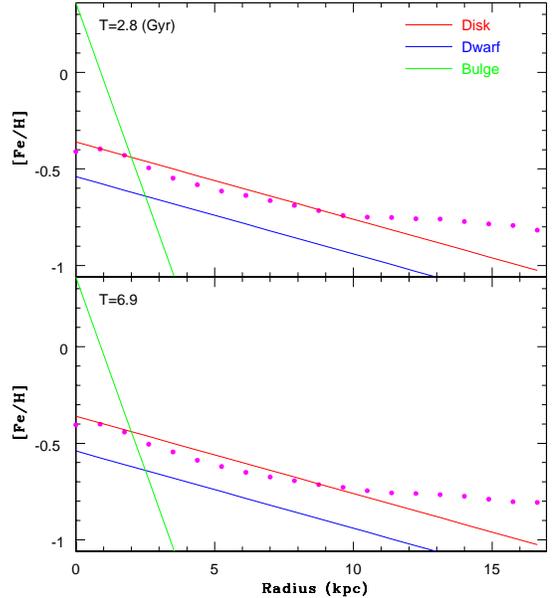}
\figcaption{
The same as Figure 5 but for the two-component disk at $T=2.8$ Gyr (upper) and
$T=6.9$ Gyr (lower) for the standard model.  The disk has a stellar bar so that
the radial metallicity gradient of the thick disk can be flattened slightly
by dynamical action of the bar (i.e.,  radial mixing of stars with different
metallicities).
\label{fig-14}}
\end{figure}

\subsection{Dynamical properties of two-component disks}

Figures 9 and 10
show the long-term  dynamical evolution of the two-component stellar disk
in the standard model (M1) in which the final mass fraction of the thick disk versus the thin
disk is  0.12.  A stellar bar can form spontaneously from global bar
instability  in the thin disk within $\sim 2$ Gyr  so that
the thin disk appears to have  a boxy-shape in the edge-on view ($T=2.8$ Gyr).
The ``box-shaped bulge", which is just an edge-on bar, appears to 
grow as the stellar bar
become stronger and longer ($T=5.5$ Gyr).
Finally a thin disk is transformed into a longer  stellar bar with the edge-on view
being similar to a disk with a boxy bulge ($T=6.9$ Gyr).
The stellar bar of the thin disk can dynamically interact with the thick disk
so that the structural and kinematical properties of the thick disk can be 
significantly changed. One of the  remarkable effects of the barred thin disk
is that the initially non-barred thick disk can be transformed into a barred
one with the position angle and pattern speed of the bar being similar to those
of the bar in the thin disk.
As shown in Figure 9,  a slightly shorter stellar bar is formed in the thick
disk at $T=2.8$ Gyr. This formation of the ``barred thick disk'' suggests that
the Galactic thick disk is likely to have a barred structure, in particular,
in its inner part. Figure 10 also shows that the thick disk appears to be  thicker
and more triaxial after its dynamical interaction with the thin disk.

Figure 11 shows the radial and vertical surface stellar density 
profiles (${\Sigma}_{\rm s}$) 
of the thin and thick disks at $T=0$ Gyr (initial) and $T=2.8$ Gyr
(after bar formation) in the standard model. This figure accordingly shows
how the radial and vertical structures of the two disks change as a result
of the dynamical interaction between the stellar bar and the two disks.
The simulated rather flat profile of the vertical structure of the thick
disk at $5 \le R \le 9$kpc in the present study can be seen also in a recent work
by Villalobos et al. (2010).
Although the inner part of the thick disk is morphologically transformed
into a bar,
the radial density profile does not change so significantly.
The central ($R<2$ kpc) and outer ($R>12$ kpc) parts  of the thin disk
show slight increases in the radial 
surface stellar density profile  (${\Sigma}_{\rm s} (R)$)
whereas the inner part at $R\sim 5$ kpc shows a decrease in ${\Sigma}_{\rm s}(R)$.
The thick disk at $5\le R \le 9$kpc does not show a significant change in
the overall vertical structure, though the profile becomes slightly steeper.  
The vertical density profile of the thin disk becomes appreciably flatter 
(or the thin disk becomes slightly thicker) owing to the dynamical action
of the bar. 
Thus, the radial and vertical density profiles of the thin disk
can be more strongly influenced by the stellar bar in comparison with those
of the thick disk.

Figure 12 shows that the thin and thick disks have different kinematical
properties just after the formation of the inner stellar bar at $T=2.8$ Gyr.
Owing to the dynamical influence of the stellar bar on the thick disk,
the $V_{\phi}$ difference in the thin and thick disk at $R=R_{\odot}$ becomes
smaller ($\sim 16$ km s$^{-1}$). The vertical velocity dispersion of the thick
disk also become slightly larger (${\sigma}_{\rm z} \sim 53$ km s$^{-1}$) after
the stellar bar formation. The thick disk has ${\sigma}_{\rm r}=70$ km s$^{-1}$
(i.e., ${\sigma}_{\rm r}/{\sigma}_{\rm z} \sim 1.3$)
and ${\sigma}_{\phi}=55$ km s$^{-1}$ (${\sigma}_{\phi}/{\sigma}_{\rm z} \sim 1.0$), 
both of which are significantly higher
than those of the thin disk.  The thick disk can continue to be dynamically
influenced by the growing stellar bar in the thin disk so that 
${\sigma}_{\phi}$ and  ${\sigma}_{\rm z}$ at $R=R_{\odot}$
can become slightly  larger at $T=6.9$ Gyr.
The derived larger  ${\sigma}_{\rm z}$ means that the stellar bar 
in the thin disk can continue 
to change the  kinematics of the thick disk within
$\sim 7$  Gyrs.

Figure 13 shows that the kinematical properties of the simulated two-component disks
depend largely 
on the model parameters of minor mergers ($m_2$, $r_{\rm p}$, and $e_{\rm p}$)
in the models, M2 $-$ M5.
The two-component disks constructed from merger remnants with $m_2=0.05$ (M2) and 0.1 (M3)
have smaller $V_{\phi}$ differences in the thin and thick disks
and smaller ${\sigma}_{\rm z}$. In these models with smaller $m_2$,
the differences in the radial profiles of $V_{\phi}$, ${\sigma}_{\rm r}$, and
${\sigma}_{\phi}$ between the thin and the thick disks
at $R<10$ kpc are small  and thus
are less consistent with observations,  which means that minor merging
with lower $m_2$ is unlikely to be  responsible for the formation of the Galactic thick
disk. The model M4 with larger $m_2$ (=0.3) shows a higher degree of kinematical
differences in the thin and thick disks, which can be still consistent with
observations. The stellar bar developed in the thick disk for this model is a shorter
(or fatter) and weaker one. The thick disk in the model M5
with $m_2=0.2$, $r_{\rm p}=0.5R_{\rm d}$,
and $e_{\rm p}=0.8$ (thus more energetic minor merger) does not clearly show
a strongly barred structure, though the thin disk has a strong bar. 
As shown in Figure 13,  the FGTD is too strongly damaged by minor merging so that
the  $V_{\phi}$ difference in the thin and thick
disk is quite large,
even after the slow growth of the thin disk. This model cannot be regarded
as reasonable  for the Galactic disk.

Although we confirm that stellar bars can be formed in thick  disks for most
models (i.e., M1-M12 except M5),  
the morphological properties of the simulated bars are different between
different models. For example, the model M12 with a smaller 
mass of the thin disk ($M_{\rm d, n}=2 \times 10^{10} M_{\odot}$)
shows the formation of a shorter and weaker
bar in the thin disk so that the thick disk cannot clearly show a strongly barred 
structure (but show a  fat, elliptic morphology)
owing to the weaker dynamical effect of the bar of the thin disk on
the thick disk. The models with smaller $M_{\rm d, n}$
(e.g., $M_{\rm d, n} \le 10^{10} M_{\odot}$) do not show the formation
of bars in the thin (thus thick) disk.
These results  suggest that the thick disk can be influenced
by the barred structure of the
thin disk only in the later stage of the disk evolution (i.e., when
the disk has a larger mass). 
These models with smaller $M_{\rm d, n}$ show lower velocity dispersion
within the thick disk (e.g., ${\sigma}_{\rm z} \sim 40$ km s$^{-1}$), which
is  more consistent with observations. The present results on the kinematical
differences between thin and thick disks do not depend so strongly
on $M_{\rm d}$, $f_{\rm dm}$, and $a_{\rm d,n}$, but they depend more strongly
on $m_2$ and ${\Theta}$. The dynamical properties
of the model M6 with large ${\Theta}$ (=60) are less consistent with 
observations than
than other models with lower $\Theta$ ($\le 30$), as already demonstrated by previous
works (e.g., Villalobos et al. 2008).

It is possible that the slow growth of the thin disk
and its  stellar bar
can also significantly change the radial
metallicity gradient owing to the  dynamical action
of the bar. Figure 14 shows that the radial metallicity gradient at $T=2.8$ Gyr
becomes  slightly flattened in comparison with the gradient  just after minor merging shown
in Figure 5.  The reason for this is that the metallicities in the inner
regions of the thick disk at $R<10$ kpc become systematically smaller after the 
formation of the stellar bar. This result implies that the growth process
of the thin disk and the dynamical action of the stellar bar can be important
factors for determining the slope of the radial metallicity gradient of the
thick disk. It should be stressed here that the flattening of the radial metallicity
gradient of the thick disk by the thin disk can be also clearly seen in the model
with lower  $m_2$ (=0.05 and 0.1): the radial metallicity gradients
of the thick disks  at $R>5$ kpc
in these two-component models become significantly more flattened than those
just after minor merging.

The dynamical interaction between stellar bars and thick disks in the present two-component
disk models can transfer stars within $R=2$ kpc to the outer disk regions at $R>R_{\odot}$.
However the number fractions of stars 
that are located initially at $R\le2$ kpc  and finally
at $R\ge 8$ kpc  is less than $10^{-4}$ for the present models, even if the central
bars  become long and strong.
Therefore, the radial transfer of bulge stars with high [Fe/H]
to the solar-neighborhood is much less
efficient in the present models.
It should be stressed here that some fraction of  metal-rich 
stars in the inner disk regions can be transferred to the solar neighborhood.
On the other hand, the slow growth of the thin disk
can contract the thick disk so that the number fractions of more metal-rich stars
([Fe/H]$>-0.5$) at $R>5$ kpc in the thick disks
can slightly decrease as the thin disk
slowly grows (though the number fractions just after minor merging
are larger).
Thus these results in \S 3.1 and 3.2 suggest that both minor merging and 
dynamical influences  of the evolving thin disk on the thick disk
can be important in 
understanding the origin of the radial metallicity gradient and the
presence of metal-rich stars at $R=R_{\odot}$ 
in the Galactic thick  disk.

\begin{figure}
\plotone{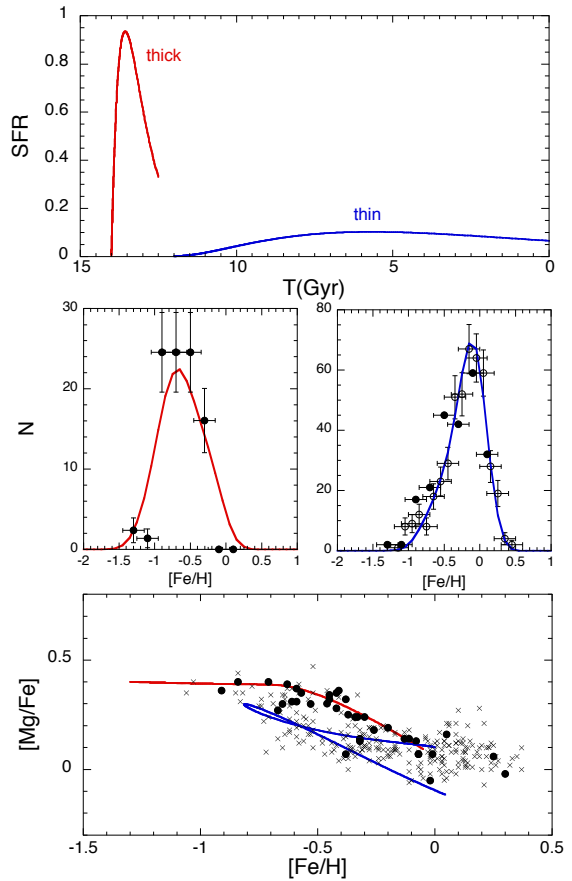}
\figcaption{
Chemical features predicted by the continuous star formation
 model for the thick and thin disks in the solar neighborhood.
{\it Upper panel}: The star formation rate as a function of time for the thick disk (red curve) and the thin disk (blue curve). 
Each SFR is normalized so that each total stellar mass is unity. 
{\it Middle panel}: Predicted MDFs of the thick disk (left) 
and the thin disk stars (right)  compared with the observations. 
The calculated distributions are
convolved using a  Gaussian with a dispersion 
of 0.1 dex considering a measurement error expected in the data. 
Filled circles represent data taken from Wyse and Gilmore (1995), 
and open circles are from Edvardsson et al. (1993). {\it Bottom panel}: 
Observed and predicted correlations of [Mg/Fe] with [Fe/H] for the thick
 and thin disks, 
compared with those for the Galaxy. The observed data of the thick disk
 are denoted by filled circles (Bensby et al.~2005), 
and those for the thin disk are denoted by crosses 
(Edvardsson et al.~1993; Bensby et al.~2005). 
\label{fig-15}}
\end{figure}

\begin{figure}
\plotone{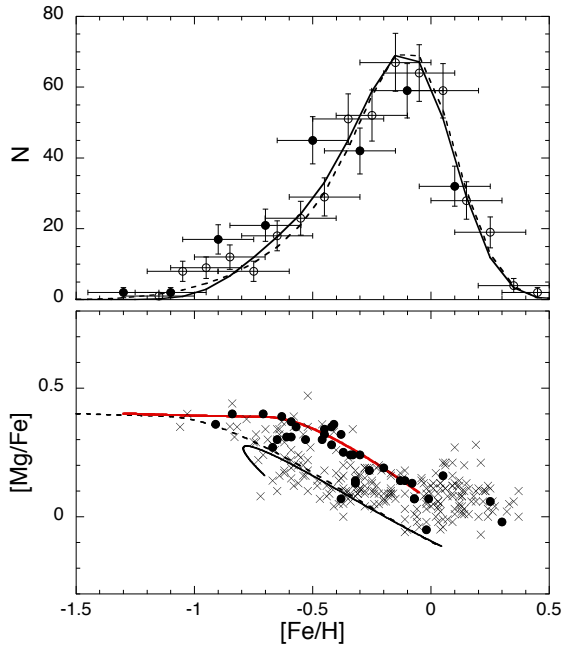}
\figcaption{
Predicted features by the temporal truncation  model (solid curve) and the expulsion model 
(dashed curve) for the thin disk. 
The symbols are the same as in Figure15. 
For reference, the result for the thick disk is shown as a  red curve in the lower panel.
\label{fig-16}}
\end{figure}

\subsection{Chemical evolution of the thin and thick disks}

Figure 15 shows the star formation histories, MDFs, 
and [Mg/Fe]-[Fe/H] relations for the thin and thick disks 
in the continuous star formation model.
The upper panel of Figure 15 shows the predicted star formation history 
of the Galactic disk, where the thick disk is first formed rapidly, 
followed by a gradual formation of the thin disk over a prolonged timescale. 
Here, each SFR is normalized so that each total stellar mass is unity. 
The resultant MDFs for thick disk and thin disk, 
and the correlations of [Mg/Fe] with [Fe/H] 
are shown in this figure.
In this model,
the thick disk formation terminates 
when the metallicity reaches [Fe/H]$\sim$0 at around 1.5 Gyr,
and accordingly 
this model explains the overall shape of the observed MDF.
However, if we assume a longer timescale of star formation,
the predicted [Mg/Fe] trend afterwards 
continues to decrease with  increasing [Fe/H]
whereas the observed [Mg/Fe] for [Fe/H]$>$0 holds broadly constant.

Therefore, the model cannot reproduce well the observed flat distribution
of the thick disk stars with [Fe/H]$>0$ in the [Mg/Fe]-[Fe/H] relation 
(e.g., Bensby et  al. 2003).
Although the total number of such stars could be less than $\sim 10$\% (e.g., Bensby et al. 2007),
this apparent inconsistency implies that such metal-rich stars
cannot form {\it in situ} at the solar-neighborhood.
Thus, we need to consider that (i) this apparent inconsistency
is due largely to the limitation of the adopted one-zone chemical evolution
model which does not consider mixing of stellar populations
between different disk regions  and thus  (ii)
these metal-rich stars with [Fe/H]$>0$ and [Mg/Fe]$ \sim0.1$
are likely to  migrate from 
the inner thick disk and the bulge where more metal-rich stars exist.
We will discuss this point later in the context of
radial  mixing due to minor merging in \S 4.2.

Since the thick disk formation leaves the metal-rich gas as an end product 
of its chemical evolution, the evolution of [Mg/Fe] starts from [Fe/H]$\sim$0
(when the final gas mass fraction is  $\sim 10$\%).
Then, [Fe/H] and [Mg/Fe] decreases  and increases, respectively,
owing to dilution by metal-poor infalling gas with a high [Mg/Fe] ratio ($\sim$+0.4).
This reverse evolution comes to an end when the chemical enrichment by star formation 
exceeds the effect of gas dilution, and subsequently an usual evolutionary 
path appears. In the end,  the  overall behavior explains, 
in part,  a large dispersion in stellar ages as well as 
in [Mg/Fe] among the thin disk stars. In addition, it may be 
possible to claim that  the remaining metal-rich 
gas after the thick disk formation results in the  
presence of no metal-poor thin disk stars.

The results of the temporal truncation model do not differ significantly from
those of the continuous star formation model  in terms of the MDF of the thin disk stars.
Figure 16 shows that the temporal truncation model can reproduce the observed
MDF of the thin disk stars as well as the continuous one. The remarkable difference
between the two is that there are no old thin disk stars with [Fe/H]$\sim 0$ and 
[Mg/Fe]$\sim 0.1$ in the temporal  truncation  model,  because star formation
in the thin disk can start only after the gas of the disk has [Fe/H]$\sim -0.7$
owing to gas accretion in the temporal truncation model.
There is a clear distinction between the locations of the thin and thick disk stars
on the [Mg/Fe]-[Fe/H] relation in the temporal truncation model versus 
the continuous  star formation model. However, it is   currently difficult 
to determine which of the two  can better fit to observations
owing to 
the lack of observational information on the relation
between [Mg/Fe], [Fe/H], and stellar ages among the thin disk stars.

The gas expulsion model shows  MDFs and [Mg/Fe]-[Fe/H] relations quite similar
to those in the other two models, which implies that it would be difficult
to observationally determine the best possible model among the three based 
only on the comparison  between the observed and simulated MDFs and 
[Mg/Fe] relations. The only significant difference between this model and other
two is the presence of thin disk stars with low [Fe/H] ($<-1$)  and high [Mg/Fe]
($\sim 0.4$). 
In the end,
the locations of the oldest thin disk stars on the [Mg/Fe]-[Fe/H] relation
can be significantly different among  the three models, which implies that
the simultaneous derivation  of ages and abundances for such older stars
is a key to distinguish between the three models.
The overall agreement of the observed
and simulated MDFs and [Mg/Fe]-[Fe/H] relations of the Galactic thin and thick
disk stars for all of the three models suggests that the minor merger
scenario of the thick disk formation is  a promising one.

\begin{figure}
\plotone{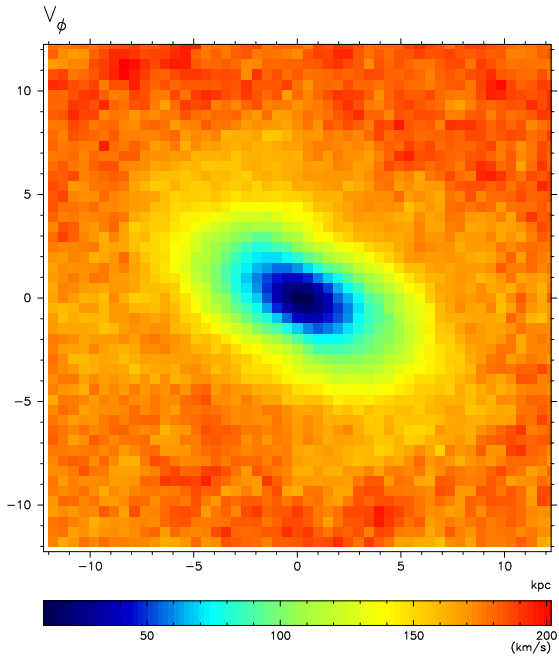}
\figcaption{
The two-dimensional (2D) distribution of  $V_{\phi}$ 
for thick disk stars at $T=2.8$ Gyr in the standard model.
The simulated 2D $V_{\phi}$ distribution is clearly asymmetric, in particular, in the central 
5kpc of the thick disk. 
\label{fig-17}}
\end{figure}

\begin{figure}
\plotone{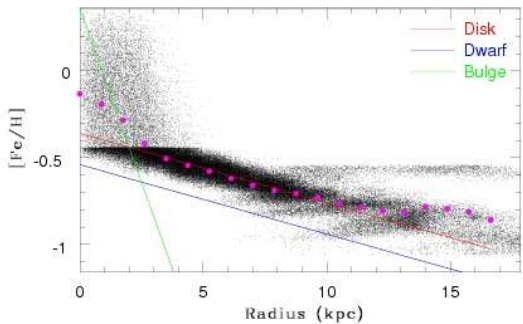}
\figcaption{
The same as Figure 5 but for the model M2 with $m_2=0.05$.
It is clear that radial mixing of stellar populations by minor merging in
this model is less efficient in comparison with the standard model M1 shown
in Figure 5.
\label{fig-18}}
\end{figure}

\section{Discussion}

\subsection{The possible presence of the barred thick disk}

The present study has first demonstrated that the Galactic thick disk can have
a barred structure
if it was formed prior to the formation of the Galactic thin disk and thus has long been
influenced dynamically by the thin disk with a stellar bar.
Given that most  models have shown the formation of the barred thick disk
in the two-component stellar disks,  the present prediction on the presence
of the barred thick disk in the Galaxy might well be  regarded as robust. 
Recent theoretical
studies on the formation of the thick disk through radial mixing of stars 
did not show the formation of the barred thick disk  in their models
(e.g., Sch\"onrich \& Binney 2009;  Loebman et al.  2010).
Previous cosmological simulations on the thick disk formation
via accretion (Abadi et al. 2003) and gas-rich mergers (Brook et al. 2004)
did not show the formation of the barred thick disk either. It is thus possible that
the presence or absence of the global barred structure of the thick disk can
be a clue as to how the thick disk was formed.
However, it should be stressed here that the present study  could overestimate 
the dynamical influence of the stellar bar on the thick disk,
because the simulated  bars are stronger and longer
than the real Galactic bar  owing to
the adopted model in which the entire disk has a smaller $Q$  ($1.5$)
when it is fully developed: a stronger and longer
stellar bar can form from  such a  massive
disk with a smaller $Q$ parameter.

Larsen \& Humphreys (1996) first revealed an  excess of faint blue stars
with $B$-$V$ bluer  than by 0.6 mag in
Quadrant 1 (Q1) in comparison with Quadrant 4 (Q4)
and thus suggested that the observed excess can result either
from the elliptic thick disk or heating of the disk by the bar.
Parker et al. (2004) later confirmed the presence of the asymmetric thick disk 
in Q1 and also revealed kinematical asymmetry between the thick disk stars in
Q1 and Q4 (i.e., differences in $V_{\phi}$). 
Humphreys et al. (2011) confirmed the kinematical differences of the thick disk
stars in Q1 and Q4 and suggested that the dynamical interaction between
the stellar bar and the thick disk can be responsible for the observed
kinematical differences. 
Given that these observations did not map the entire region of the thick disk,
it remains observationally unclear 
whether or not  the thick disk stars have a global barred structure, in particular,
in the inner disk ($R<3$ kpc), where the Galactic bar is considered to exist
(e.g., Weinberg 1992).

Although the above observations cannot be compared directly with the present results,
it is instructive for the present study to show how $V_{\phi}$ in the thick disk
can be different in different regions of the Galaxy.
Fig. 17 shows the 2D distribution of $V_{\phi}$ for  thick disk stars projected
onto the $x$-$y$ plane  at $T=2.8$ Gyr
in the standard model.  The simulated thick disk with a central bar 
in this model is divided into
$50 \times 50$ local regions within 12 kpc from the center of the disk
and the mean $V_{\phi}$ at each region is estimated.
The derived 2D $V_{\phi}$ distribution clearly shows an elongated and asymmetric (bar-like)
feature, in particular, for the inner region ($R<5$ kpc).  
This result clearly suggests  that
if observational studies find an asymmetric 2D
$V_{\phi}$ distribution in the thick disk,
there is  kinematical evidence which  supports
the presence of  a barred thick disk in the Galaxy.
There is no clear sign
of a past minor merger event in the 2D $V_{\phi}$ distribution,
which implies that kinematical evidence for past minor merger events
would be hard to find in observations on the 2D $V_{\phi}$ distribution
of the Galactic thick disk.

If the Galactic thick disk really has a barred structure in its inner part,
then the observed 2D structure and kinematics can give strong constraints
on the formation processes of the thick disk. 
A discovery of a barred structure with the position angle and pattern speed
of the bar being almost identical to those of the Galactic bar would support
the minor merger scenario.  If such a bar is not found in the inner region
of the Galaxy,  then the minor merger scenario would need to clarify
why the formation of a bar in the thick disk is suppressed in spite of the
presence of ongoing interaction between  the realistic bar and  the thick disk.
Given that there is no paper that extensively  investigated dynamical interaction
between the realistic Galactic bar and the thick disk,
it is doubtlessly worthwhile for future studies
to investigate their  dynamical interaction
in a comprehensive way,
and compare the simulated 2D radial and azimuthal velocity dispersions in the
thick disk with the corresponding observations (e.g., Humphreys et al. 2011).

\subsection{The radial  metallicity gradient in the thick disk}

Allende Prieto et al. (2006) revealed that G-dwarf stars 
in the Galactic thick disk have 
a very flat radial metallicity gradient for $4<R<14$ kpc and $1<|z|<3$ kpc
and that the mean metallicity at each radial bin for $4<R<14$ kpc
is about [Fe/H]$\sim -0.7$ (independent of $R$). Although the presence of
more  metal-rich stars in the outer part ($R>10$ kpc) of the thick disk
is consistent with the present merger scenario, 
the observed no/little radial metallicity gradient appears to be less consistent 
with the predictions of the present minor merger models. 
The above mentioned problem of the present 
minor merger scenario is due to the larger mean metallicity ([Fe/H]$>-0.7$)
of the thick disk stars at $4<R<7$ kpc. For the minor merger scenario to be
still a promising one,  the problem would need to be resolved. 
Obviously, if the original metallicity gradient in the FGTD is flat, then
the final one in the thick disk can be also flat. If this is the case,
the minor merger scenario has no serious problem with the observed lack of a
metallicity gradient in  the thick disk.
However, star formation can proceed more rapidly in higher density regions so that
chemical abundances become higher more rapidly in these  regions.
Therefore it is less likely  that the FGTD with an initial  exponential radial density
profile (i.e., higher density in the inner region)  has no 
radial metallicity  gradient. Thus the above apparent inconsistency on the
radial metallicity gradient  
would need to be regarded
as a potential  problem in the minor merger scenario.

The observed lack of the radial metallicity gradient in the thick disk
is suggested to be consistent with the gas-rich merger scenario for  thick
disk formation (e.g., Brook et al. 2004; Allende Prieto et al. 2006). 
Although the radial mixing scenario by 
Sch\"onrich \& Binney (2009) did not extensively discuss
the radial metallicity gradient of the thick disk, 
Fig. 12 in Loebman et al. (2010)  shows a flat
(or possibly positive) gradient
$7<R<11$ kpc, though the thick
disk stars appears to have  higher [Fe/H] ranging from $-0.5$ to $-0.3$
in their Figure 6.
Since numerical simulations 
based on the accretion scenario by Abadi et al. (2003) did not include
chemical evolution models,  we cannot discuss the scenario  here in  the context
of the radial metallicity gradient of the thick disk.

Radial mixing of stellar populations 
during minor merging can transfer stars with higher metallicities
in the inner  FGTD ($R<5$ kpc) 
to the solar neighborhood ($R=R_{\odot}$, 8.5kpc)  and thus
provide a clue to the origin of  the observed presence of metal-rich
stars with [Fe/H]$ > 0$ and higher [$\alpha$/Fe] ($ \sim 0.1$) in the thick
disk (e.g., Bensby et al. 2003; Casagrande et al. 2011).
In the present study, the final number fractions ($F_{\rm mr}$)
of  metal-rich stars with
[Fe/H]$>0$  at $R>8$ kpc in the two-component
models are  less than $10^{-4}$ (even after the formation
of bars). This number $F_{\rm mr}$ can increase significantly
if we adopt a more realistic model in which
the initial MDF at each radius in the FGTD
has a dispersion: the metallicity at each radius in the FGTD is fixed at
a certain value (with no dispersion) 
according to the adopted metallicity gradient
in the present study.
It should also be  stressed here that minor merging with $m_2 \le 0.05$ does  not
show such efficient radial mixing as seen in other models with $m_2 \ge 0.1$
in the present study. Figure 18 shows that (i) the metallicity gradient
becomes flat only for $R>10$ kpc in the model M2 with $m_2=0.05$ and
(ii) the transfer of the metal-rich  stars initially  within $R<5$ kpc
to the solar neighborhood is much less efficient. 
The radial mixing scenario (e.g., Ro\v skar et al. 2008; Loebman et al. 2010)
can explain the presence of such metal-rich stars in a different way
as the present merger scenario.

\begin{figure}
\plotone{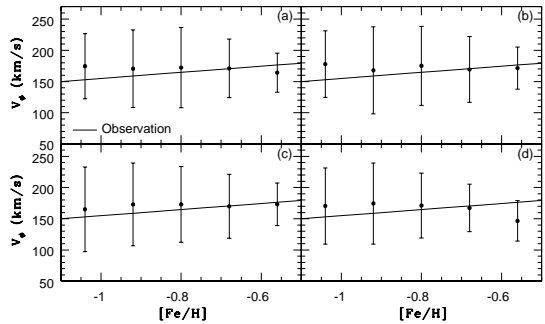}
\figcaption{
Dependence of $V_{\phi}$ 
for  thick disk stars at the solar-neighborhood on [Fe/H] in
the standard model with different metallicity gradients for the FGTD:
(a) ${\alpha}_{\rm d}=-0.04$,
(b) ${\alpha}_{\rm d}=-0.06$,
(c) ${\alpha}_{\rm d}=-0.08$,
and (d) ${\alpha}_{\rm d}=-0.04$ and the central metallicity of the FGTD 0.1 dex lower
than that in the model (a). 
Each filled circle
and error bar 
represent the mean $V_{\phi}$ and its dispersion at each [Fe/H] bin.
The observed positive $V_{\phi}$-[Fe/H] correlation from Lee et al. (2011) 
is shown by a solid line in each frame.
Here stars with $8\le R \le 9$
kpc and $1 \le |z| \le 3$ kpc are selected so that the simulation results  can be 
compared with the corresponding observations.
\label{fig-19}}
\end{figure}

\begin{figure}
\plotone{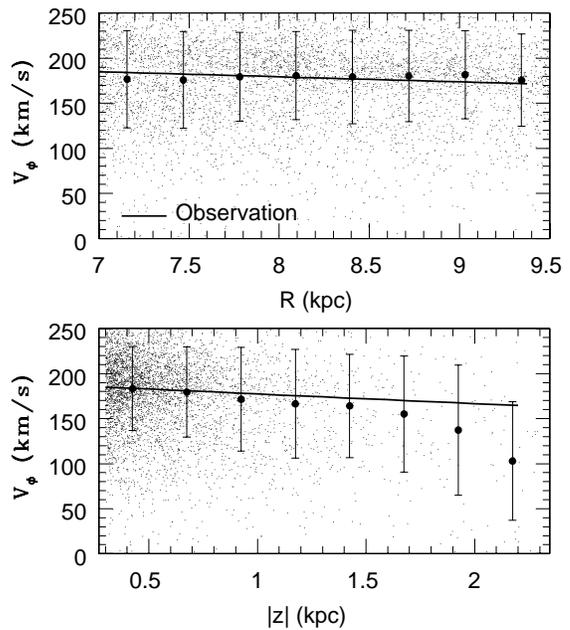}
\figcaption{
Dependence of $V_{\phi}$ on $R$ (upper) and $|z|$ (lower) in the model M15 which can
reproduce well the observations  by Lee et al. (2011). The mean $V{\phi}$
and dispersion in $V_{\phi}$ at each bin are shown by filled circles and error bars,
respectively. The small dots show the results of individual thick disk particles in the model.
The observed correlations from Lee et al. (2011) are shown by solid lines.
\label{fig-21}}
\end{figure}

\subsection{The observed $V_{\phi}$-[Fe/H] relation 
of the thick disk}

Chiba \& Beers (2000) investigated space motions and metal-abundances of
1203 solar-neighborhood stars with [Fe/H]$< -0.6$ and found that $V_{\phi}$
of the more metal-rich stars are larger for $|z|<4$ kpc (see also more recent
results by Carollo et al. 2010). Spagna et al. (2010) also found that 
more metal-rich stars in the Galactic thick disk at $1<|z|<3$ kpc have
higher $V_{\phi}$ and suggested that this correlation can give a constraint
on the formation models of the thick disk. 
Lee et al. (2011) have recently confirmed a positive correlation between $V_{\phi}$ and
[Fe/H] in the thick disk and derived a relation of $\Delta V_{\phi}/\Delta$[Fe/H]=49.0$\pm 3.4$
(km s$^{-1}$ dex$^{-1}$).
If future observations (e.g., GAIA and HERMES) can select  thick disk stars
based both on their locations and kinematics and thereby investigate 
the $V_{\phi}$-[Fe/H] relation, then we will be able to
make a more robust conclusion on the viability of the minor merger scenario 
based on the comparison between the observed and simulated
$V_{\phi}$-[Fe/H] relations.
However, it is  worthwhile  for the present study to investigate whether there is 
a positive correlation between [Fe/H] and $V_{\phi}$ in the simulated thick disks.

Figure 19 shows  the $V_{\phi}$-[Fe/H] relation for 1789 stars of the thick
disk at $8 \le R \le 9$ kpc and $1 \le |z| \le 3$ kpc  for the two component
disk of the standard model (at $T=2.8$ Gyr).
Here four different models for different initial metallicity gradients in the outer
part of the FGTD 
are adopted.
Although the dispersion at each [Fe/H] bin is large ($\sim 50$ km s$^{-1}$),
there is no clear  $V_{\phi}$-[Fe/H] correlation for ${\alpha}_{\rm d}=-0.04$
in the standard model.
In the present minor merger scenario,  more metal-rich stars are initially
located in the inner regions of the FGTD (owing to the adopted
negative metallicity gradient of the FGTD), where $V_{\phi}$ is smaller,
so that they can have smaller $V_{\phi}$. Therefore, it is not so straightforward
for the minor merger scenario to explain the higher $V_{\phi}$ in more metal-rich
stars at $R=R_{\odot}$.  
It is found that for a steeper metallicity gradient (${\alpha}_{\rm d}=-0.08$),
the simulated $V_{\phi}-$[Fe/H] relation becomes similar to  the observed
one by Lee et al. (2011), which shows a weak positive correlation between $V_{\phi}$ and
[Fe/H]. 
There would be a possibility that the thin disk stars with higher [Fe/H] and larger 
$V_{\phi}$ are included in the observed thick disk sample. If this
contamination of the thin disk stars is real, then the observed $V_{\phi}-$[Fe/H]
becomes even weaker so that it can be more consistent with the simulated one in the present
study.
We confirm that models with ${\alpha}_{\rm d}=-0.04$ and different central
metallicities in the disk (e.g., (d) in Figure 19) cannot reproduce the observation very well.
These results imply that
the observed  correlation between $V_{\phi}$ and [Fe/H] 
can give some constraints on the initial metallicity gradient of the FGTD in the Galaxy.

\begin{figure}
\plotone{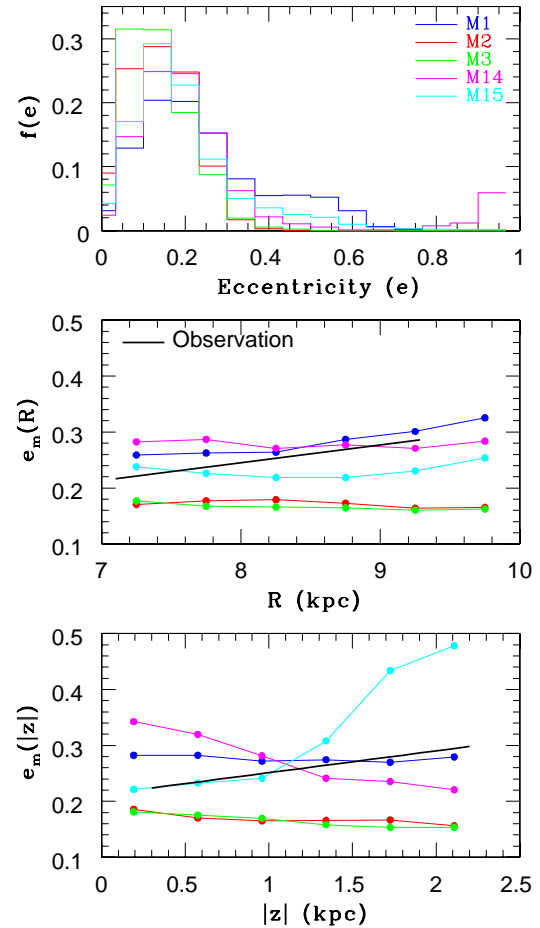}
\figcaption{
Distributions of orbital eccentricities ($f(e)$, top),  dependence
of mean orbital eccentricities ($e_{\rm m}$) 
on  $R$ (middle), and those on 
$|z|$ (bottom), for five different models:
the standard M1 (blue), M2 (red), M3 (green), M14 (magenta),  and M15 (cyan).
The observed correlations from Lee et al. (2011) are shown by solid lines
in the middle and bottom panels.
\label{fig-21}}
\end{figure}

\begin{figure}
\plotone{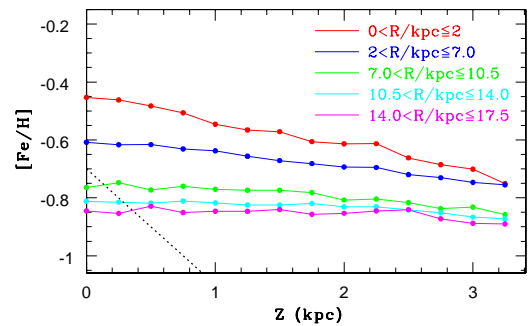}
\figcaption{
Vertical metallicity gradients at different radii in the standard model:
at $0 <R \le 2$ kpc (red), 
$2 <R \le 7.0$ kpc (blue), 
$7.0 <R \le 10.5$ kpc (green), 
$10.5 <R \le 14$ kpc (cyan), 
and $14.0 <R \le 17.5$ kpc (magenta).
For comparison, the original vertical metallicity gradient of the FGTD at $R=8.5$ kpc is shown
by a thick dotted line.
\label{fig-22}}
\end{figure}

\subsection{$V_{\phi}$ and  orbital eccentricities
of the thick disk stars   dependent on $|z|$}

Recent observational studies have revealed that 
$V_{\phi}$ depends on  $|z|$  such that $V_{\phi}$ is smaller for larger $|z|$ 
($\Delta V_{\phi}/ \Delta |z| = -10.8 \pm0.9$ km s$^{-1}$ kpc$^{-1}$)
for the Galactic thick disk at the solar neighborhood
(Lee et al. 2011).
Although Villalobos \& Helmi (2008) have already shown that thick disks 
which formed from
minor merging with lower $\Theta$ ($0$ and $30$ in their Figure 20) can clearly show
$V_{\phi}-|z|$ correlations, they did not directly compare their results with the corresponding
observations. Accordingly, it is worthwhile for the present study to compare
the results with the observational result by Lee et al. (2011).
We have investigated which model in the present study can reproduce the observed $V_{\phi}-|z|$
and  $V_{\phi}-R$ 
relations best among the representative models and found that  model M15 can show 
these relations similar to the observed ones better than other models
(including M16 with $m_2=0.2$ and $\Theta=0$). Figure 20 shows
the results of  model M15 in which $\Theta =0$ and $m_2=0.1$ are adopted.
The observed $V_{\phi}-|z|$
correlation can be better reproduced in models with low $\Theta$ ($\le 30$) in the present study,
which is consistent with the results by Villalobos \& Helmi (2008).
The models with low $\Theta$ show no strong positive correlation between $V_{\phi}$ and $R$ whereas
observations (Lee et al. 2011) show a rather weak correlation 
($\Delta V_{\phi}/ \Delta R = -5.6 \pm1.3$ km s$^{-1}$ kpc$^{-1}$).

Lee et al. (2011) furthermore have revealed that
the orbital eccentricities of thick
disk stars at the solar neighborhood show a correlation between $e$ and $|z|$ (and $R$).
If the thick disk formed before the formation of the thin disk in the Galaxy,
then orbital eccentricities of the thick disk stars could be influenced by the formation
processes of the thin disk. Previous simulations on the eccentricity distribution ($f(e)$)
of the thick disk stars however did not consider the evolution of $f(e)$ 
during the formation of the thin disk. We accordingly have investigated $f(e)$ based on 
the first $\sim 3$ Gyr
orbital evolution  of the simulated thick disk
stars  for different  models (after the formation of
the thin disk yet before the strong influence of the stellar bar formed
in the thin disk). For the purpose of direct comparison between
simulations and  observations from Lee et al. (2011),
the thick disk stars with $7 \le R \le 10$ kpc and $0 \le |z| \le 2.3$ kpc are selected to derive
$f(e)$ and mean orbital eccentricity ($e_{\rm m}$) 
at a given $R$ and $|z|$ in each  model.

Observations on $e_{\rm m}$ at different radial and vertical bins  by Lee et al. (2011)
show positive correlations of $e_{\rm m}$ with  $R$  and $|z|$
(i.e., larger $e_{\rm m}$ for larger $R$ and $|z|$).
Figure 21 shows $f(e)$ and $e_{\rm m}$ dependent  on $|z|$ (and $R$) derived in the five
representative models and accordingly can be compared with observational results shown
in Figures 9 and 10 by Lee et al.
(2011).
The standard model M1 with $e_{\rm p}=0.5$
shows an orbital eccentricity distribution similar to the observed one
(Lee et al. 2011) whereas the models with low $m_2$ (0.05 and 0.1)
and $e_{\rm p}=0.5$ show narrower distributions with
their peaks at lower $e$. The model M14 with $m_2=0.05$ and $e_{\rm p}=0.8$ shows
a small bump at higher $e$ ($>0.8$), which is not observed in  Lee et al. (2011).
Furthermore the dependency of $e_{\rm m}$ on $|z|$ in this model M14
is qualitatively inconsistent
with the observed one. 
The model M15 with $m_2=0.1$, $e_{\rm p}=0.6$, and $\Theta=0$, which can explain
better the observed $V_{\phi}-|z|$ relation (in Figure 20), shows a peak around $e\sim 0.15$,
which is slightly smaller than the observed peak around $e \sim 0.2$. 
These results imply that the models with larger $m_2$ ($\sim 0.2$)
and lower $e_{\rm p}$ ($\sim 0.5$) can better reproduce the observed $f(e)$ in the Galactic
thick disk.

The standard model M1 shows a very weak positive correlation between
$e_{\rm m}$ and $R$ (i.e., larger $e_{\rm m}$ at larger $R$) whereas models with lower $m_2$
show rather flat $e_{\rm m}-R$ relations. 
The observations by Lee et al. (2011) show a weak positive flat 
$e_{\rm m}-R$ relation (i.e., larger $e_{\rm m}$ for larger $R$),
which appears to be more consistent with models with higher $m_2$.
These results imply that the observed $e_{\rm m}-R$ relation
can give some constraints on $m_2$ of minor mergers that formed the thick disk. 
Only the model M15 (lower $\Theta$) clearly shows a positive correlation
between $e_{\rm m}$ and $|z|$ (i.e., larger $e_{\rm m}$ for larger $|z|$),
which appears to be more consistent with the observation by Lee et al. (2011),
though
other models do not show such a correlation. 
This result for model M15 is consistent with those in Figure 20 
which show that $V_{\phi}$ is lower
at lower $|z|$: stars with lower $|z|$ have lower $e$ 
(i.e., more circular orbits) and thus higher $V_{\phi}$. 
The observed $e_{\rm m}-|z|$ relation combined with the $V_{\phi}-|z|$ relation
imply  that if the thick disk was formed by minor merging, then
the orbital inclination angle ($\Theta$) of merging dwarf disks should be low.
Thus the observed $f(e)$, $e_{\rm m}-|R|$  correlation,  and $e_{\rm m}-|z|$ 
can give some constraints on the present  formation model of the Galactic thick disk.

\subsection{Vertical metallicity gradient in the thick disk} 

Recently Allende Prieto et al. (2006) have revealed that there is no vertical metallicity
gradient discernible in the thick disk between $1<z<3$ kpc. Karata\c s and Klement (2011)
did not find a clear vertical abundance ([M/H]) gradient in the thick disk either.   
Although there are only a few observational studies which investigate the vertical metallicity
gradient of  the thick disk stars,
the observed gradient can be used for distinguishing  between different theoretical models
for the thick disk formation. 
We accordingly have investigated  the final vertical metallicity gradient of
the thick disk in the standard two-component disk model
based on a model for the initial vertical metallicity gradient of the FGTD.
The initial metallicity
of a star with $r=R$ and  a vertical distance ($|z|$) from the $x$-$y$
(equatorial) plane  of the FGTD 
depends also on $|z|$ and
it is assigned as follows:
\begin{equation}
{\rm [m/H]}_{\rm r=R, |z|} = {\rm [m/H]}_{|z|=0} + {\alpha}_{\rm v} \times |z|,
\end{equation}
where ${\rm [m/H]}_{\rm r=R, |z|=0}$ is a metallicity at $r=R$ and $|z|=0$.
We adopted  the slope ${\alpha}_{\rm v} \approx -0.4$ that is consistent with observations
(Frogel et al. 2000). The metallicity of a star is therefore determined
by the  equations (2), (3), and (5) according to its positions $R$ and $|z|$.
It should be stressed here that  equation (5) is used {\it only} for investigating  the vertical
gradient shown in Figure 22: other results shown in Figures 1-21   
are not based on the equation (5).
During minor merging and evolution of the two-component disk,
the initial vertical metallicity gradient of the FGTD can change owing to dynamical influences
of minor mergers  and the stellar bar.

Figure 22 shows vertical metallicity gradients of the thick disk at different radii ($R$)  at
$T=2.8$ Gyr in the standard model. It is clear that irrespective of $R$,
the original steep gradients can be
significantly flattened. This is mainly because minor merging can cause vertical
heating of the initially thin disk (the FGTD)  and thus mixing of stellar populations
with initially different metallicities. 
The vertical metallicity gradient at the solar neighborhood ($7 <R \le 10.5$ kpc) in the model
shows a rather flat (or a very  weakly  negative) gradient, which is consistent with
the above-mentioned two observations. This rather flat gradient is however not so consistent
with observations by Ivezi\'c et al. (2008), which show a negative vertical metallicity gradient
for  stars at $|z|>$ 1 kpc.
The vertical metallicity gradients are steeper
in the inner regions ($R<7$ kpc) of the simulated thick disk in comparison with
the outer ones,  because the vertical heating is more efficient in
the outer part of the stellar disk.
The derived rather flat vertical gradient at $R>7$ kpc can be seen
in other representative models of the present study. 
Given that observational studies have not yet extensively investigated
the vertical metallicity gradient of the thick disk,
it would be fair for us to claim that if future
observations confirm  
the lack of a strongly negative vertical metallicity gradient in the thick disk,
then the present minor merger scenario of the thick disk formation is supported.

\section{Conclusions}

We have investigated chemical and dynamical properties of the Galactic thick disk
formed by ancient minor  merging
based on the results of N-body numerical simulations and chemical evolution models. 
In the present minor merger scenario,  dynamical evolution of the thick disk
can be influenced by the mass growth and non-axisymmetric structures of the thin
disk whereas the early star formation history and chemical evolution of the
thin disk can be influenced both by gas left behind from the thick disk formation
(i.e., the FGTD) and by dynamical properties of the thick disk.
The main results are summarized as follows:

(1) Minor merging can cause radial mixing of stars with different metallicities
so that more metal-rich stars initially in the inner part of the FGTD
can be transported into the outer region. As a result of this,
the thick disk formed
by minor merging can have stars with [Fe/H]$\sim -0.3$ at $R=0.5R_{\rm d}$ 
(corresponding roughly to $R=R_{\odot}$)
and those with [Fe/H]$\approx -0.7$
at $R=R_{\rm d}$ just after merging,
even if such  metal-rich stars  do not exist there initially.
Therefore 
the mean metallicities in the outer regions ($R>0.5R_{\rm d}$)
of the disk can increase significantly  owing
to this mixing. 
This radial mixing of different stellar 
populations during minor merging is more effective in mergers
with larger  $m_2$.
Thus the presence of more metal-rich stars in the outer part of the Galactic
thick disk
can be  possible evidence for  ancient minor mergers  occurring  in 
the early dynamical history of the Galaxy.

(2) As a result of radial mixing of stars during minor merging,  
the original metallicity
gradient of the FGTD can become significantly flattened in the final thick disk.
Although the final
radial metallicity gradients of the simulated thick disks depend on model parameters
of the original metallicity gradients,  the present models do not show
the very flat radial metallicity gradient
that was derived in some previous observations.  The simulated metallicity gradients 
are more flattened in the outer part ($R>0.5R_{\rm d}$) for most models
in the present study. Radial mixing can also cause the differences in MDFs between
different regions of the thick disks.

(3) The pre-existing thick disk can be strongly influenced by the growth processes
of the thin disk, in particular, by the formation of a stellar bar in the
thin disk.  The thick disk can be transformed into a ``barred thick disk''
with  position angle and  bar length similar to those of the stellar bar
in the thin disk. Given that the formation of the barred thick disk
can be seen in most models of the present study, we suggest that
the bar formation in the thick disk is highly likely  if the thick disk was formed
prior to the thin disk formation. 
In some models,  the inner part of the thick disk can be vertically expanded
due to the dynamical interaction with the thin disk.
The barred thick disk cannot be formed in models with lower masses of the thin disks
($M_{\rm d, n} \approx 10^{10} M_{\odot}$)
owing to the absence of the strong stellar bars in the thin disks.

(4) The final structural and kinematical properties of the thick disk after
the formation of the thin disk depend largely on $m_2$. 
Although minor mergers with $m_2=0.2$ can explain better the observed differences
in $V_{\phi}$, ${\sigma}_{\rm r}$, ${\sigma}_{\phi}$, and ${\sigma}_{\rm z}$
between the thin and thick disks at least qualitatively, 
the simulated ${\sigma}_{\rm z}$ at $R=R_{\odot}$ 
can be slightly  larger than the observed one.
Most models  with
smaller $m_{2}$ ($=0.05$)  are difficult to 
explain so well the observed difference in $V_{\phi}$ 
at $R=R_{\odot}$ between the thin and thick disks. 
The observed kinematical differences between the thin and thick disks 
can give some constraints on physical parameters of minor merging, if the thick
disk was indeed formed by minor merging.

(5) The final thick disks in most models
do not clearly show  positive correlations between $V_{\phi}$ and [Fe/H]
at $8 \le R \le 9$ kpc (i.e., $R \sim R_{\odot}$) and $1 \le |z| \le 3$ kpc.
The models with steeper metallicity gradients with ${\alpha}_{\rm d} \sim -0.08$
for the FGTD can however show a weak yet positive $V_{\phi}$-[Fe/H] correlation. 
Some observations clearly show a positive $V_{\phi}$-[Fe/H] correlation
(i.e., higher $V_{\phi}$ for more metal-rich stars), which  suggests
that the observed $V_{\phi}$-[Fe/H] correlation can give strong constraints
on the present minor merger model, in particular, the radial metallicity gradient
of the FGTD. 
However,
it would be  possible that the   stars used for observational derivation of the 
$V_{\phi}$-[Fe/H] relation include  thin disk stars which  formed later 
in the Galactic disk.

(6) The present numerical studies have confirmed that
only models with lower $\Theta$ ($\le 30$) can reproduce the observed $V_{\phi}-|z|$ correlation
in the Galactic thick disk. 
The simulated $V_{\phi}-R$ relations in most models are rather flat whereas the latest observations
show a very weakly negative correlation between  $V_{\phi}$ and $R$ 
(smaller $V_{\phi}$ for larger $R$) for the thick disk. This implies that
the present minor merger model needs to improve in terms of reproducing the observed
$V_{\phi}-R$ correlation.
Accordingly  the observed  $V_{\phi}-|z|$
and $V_{\phi}-R$ correlations can be used for determining model parameters that
are the most reasonable for the thick disk formation.

(7) The simulated orbital eccentricity distribution ($f(e)$) of the thick disk stars
in the solar neighborhood is  in good agreement with the observations
for the models with higher $m_2$ ($\sim 0.2$) and lower $e_{\rm p}$ $(\sim 0.5)$. 
The models with higher $e_{\rm p}$ show bumps at higher $e$ ($>0.8$) in the simulated $f(e)$,
which is inconsistent with observations. The observed mean orbital eccentricities ($e_{\rm m}$)
of the thick disk stars
dependent on $R$ can be reproduced by some models in the present study.
Although the latest observations have revealed 
a positive $e_{\rm m}-|z|$ correlation (larger $e_{\rm m}$ for higher $|z|$)
in the Galactic thick disk,
only some models with low $\Theta$ can clearly show such a positive correlation
in the present study.

(8) The star formation timescale of the FGTD needs to be $\sim 1.5$  Gyrs so that
the observed MDF and the [Mg/Fe]-[Fe/H] relation of the thick
disk   can be simultaneously
explained.  Our one-zone chemical evolution models cannot explain the observed
metal-rich 
stars with [Fe/H]$>0$ and [Mg/Fe]$>0$  at $R=R_{\odot}$ owing to the
the model assumption that no radial mixing can occur.
As suggested by the present N-body simulations,  the observed metal-rich stars
in the thick disks at $R=R_{\odot}$ can be transferred from the inner
regions of  the FGTD with
the higher metallicity owing to radial mixing during minor merging.

(9) The chemical evolution of the thin disk 
depends on how star formation proceeds from
the infalling halo gas mixed with
the remaining gas left behind from the formation of the thick disk.
If the remaining gas of the thick disk can be used for
star formation in the thin disk at the early stage of the thin disk evolution,
then the new stars can have high [Fe/H] ($\sim 0$)
and low [$\alpha$/Fe] ($\sim 0.1$). 
Alternatively,
if the remaining gas of the thick disk is  expelled from
the disk region  through galactic wind (e.g., due to starbursts triggered
by minor merging),
the first stars of the thin disk have lower
[Fe/H] and higher [Mg/Fe].  
Such delayed star formation would be  also possible
if the remaining gas of the thick disk cannot be  converted into stars
until enough  halo gas can be accreted onto the disk
to effectively dilute the remaining gas of the thick disk.
Although the details of the evolution of the thin disk on the [Mg/Fe]-[Fe/H]
plane  depend
on how the remaining  gas of the thick disk mixes with the infalling halo gas, 
the present chemical evolution models can reproduce reasonably well the observed MDF and
[Mg/Fe]-[Fe/H] relation of the thin disk in a self-consistent manner.

(10) Our chemical and dynamical models suggest that the co-evolution of the 
Galactic thin and thick disks is quite important for better understanding the
origin of the two distinctive Galactic disks. 
As demonstrated by extensive comparison between the observed and simulated
properties of the thick disk,  the observed correlations between $V_{\phi}$,
$R$, $|z|$ and [Fe/H] of the thick disk stars can give strong constraints
on model parameters in the formation scenarios of the thick disk. 
The present study
did not extensively discuss the physical properties of the thin disk and the bulge
of the Galaxy.
We therefore plan to explore a theoretical model that 
can reproduce  physical properties of the thin and thick disks
and the bulge fully self-consistently  
in our forthcoming papers.


\acknowledgments
We are  grateful to the referee Timothy Beers for his constructive and
useful comments that improved this paper.
We are grateful also to Jonathan Diaz, who reviewed this paper and gave us
scientific comments and English corrections of the paper.
KB acknowledges the financial support of the Australian Research Council
throughout the course of this work.

\vspace{10cm}

\appendix
\section{Possible suppression of star formation in the early evolution
phase of the thin disk by the thick disk ?} 

Although the temporal truncation model for the chemical evolution  of the 
Galactic thin disk
has provided some explanations for the origin of the observed [Mg/Fe]-[Fe/H]
relation of the thin disk, 

it did not show how the early star formation of the thin disk
can be truncated until the gas metallicity of the thin disk becomes as low as
[Fe/H] $\sim -0.7$.
Accordingly the temporal truncation model  needs to explain why such severe
suppression of star formation could occur in the early formation history
of the thin disk. One of possible physical reasons for the suppression
of star formation in the thin disk is a higher Toomre's $Q$ parameter 
of the thin disk due to the higher stellar velocity dispersion of the
thick disk. Wang \& Silk (1994) proposed that one of important
parameters for galactic star formation in the two-component
disk composed of gas and stars 
 is the effective $Q$ parameter ($Q_{\rm e}$),
which is defined as $Q_{\rm e}^{-1}=Q_{\rm s}^{-1} + Q_{\rm g}^{-1}$,
where $Q_{\rm s}^{-1}$ and $Q_{\rm g}^{-1}$ are the stellar and gaseous
$Q$ parameters, respectively. Given that the radial velocity dispersion    
of the thick disk can be  a factor of $\sim 2$ higher than 
that of the initial the FGTD  depending
on $m_2$
(thus $Q_{\rm s}$ becomes larger than $\sim 3$), 
the two-component Galactic disk just after the thick
disk formation might well have a higher $Q_{\rm e}$.

If we adopt $Q_{\rm s}=3$ and if we assume that star formation in the thin disk
can happen for $Q_{\rm e}<1$, then $Q_{\rm g}$ needs to be less than 1.5. 
Therefore, it is possible that star formation in the thin disk cannot efficiently
occur until $Q_{\rm g}$ becomes  as low as 1.5 owing 
to the increase of the surface gas density (and/or the decrease of the 
gaseous velocity dispersion). 
The possible higher $Q_{\rm e}$ in the early stage of the thin disk formation
therefore would be responsible for  severe suppression of star formation
in the thick disk for the minor merger scenario.
If all of the remaining gas were expelled
from the thick disk
after the minor merger owing to energetic stellar winds associated
with a triggered starburst, then new stars in the thin disk can be formed
purely from gas accreted onto the disk from the Galactic halo.
If this is the case, then there is no chemical connection between the thin
and thick disks, and the above severe suppression of star formation
in the thin disk does not need to be considered in the minor merger scenario.

\newpage


\end{document}